\documentclass[aps,twocolumn,groupedaddress]{revtex4}

\usepackage{graphicx}

\newcommand{\beq}{\begin{equation}}
\newcommand{\eeq}{\end{equation}}
\newcommand{\bea}{\begin{eqnarray}}
\newcommand{\eea}{\end{eqnarray}}
\newcommand{\ba}{\begin{array}}
\newcommand{\ea}{\end{array}}
\newcommand{\bc}{\begin{center}}
\newcommand{\ec}{\end{center}}
\newcommand{\lsimeq}{\stackrel{<}{\scriptstyle\sim}}
\newcommand{\gsimeq}{\stackrel{>}{\scriptstyle\sim}}

\newcommand{\bml}{\begin{mathletters}}
\newcommand{\eml}{\end{mathletters}}
\newcommand{\commentout}[1]{{}}
\newcommand{\kvec}{{\bf k}}

\newcommand{\q}{{\bf q}}
\newcommand{\p}{{\bf p}}
\newcommand{\rvec}{{\bf r}}

\newcommand{\half}{\hbox{$1\over2$}}
\newcommand{\quarter}{\hbox{$1\over4$}}
\newcommand{\db}{{\bf d}}
\newcommand{\E}{{\bf E}}
\newcommand{\eq}[1]{(\ref{#1})}

\begin{document}
\bibliographystyle{apsrev}


\title{Theory of coherent photoassociation of a Bose-Einstein condensate}



\author{Marijan Ko\u{s}trun}
\author{Matt Mackie}
\author{Robin C\^{o}t\'e}
\author{Juha Javanainen}
\affiliation{Department of Physics, University of Connecticut, Storrs,
Connecticut 06269-3046}


\date{\today}

\begin{abstract}
We study coherent photoassociation, phenomena analogous to coherent
optical transients in few-level systems, which may take place in
photoassociation of an atomic Bose-Einstein condensate but not in a
nondegenerate gas. We develop a second-quantized Hamiltonian to describe
photoassociation, and apply the Hamiltonian both in the momentum
representation and in the position representation (field theory). Solution
of the two-mode problem including only one mode each for the atomic and
molecular condensates displays analogs of Rabi oscillations and rapid
adiabatic passage. A classical version of the field theory for atoms and
molecules is used to demonstrate that, in the presence of photoassociating
light, a joint-atom molecule is unstable against growth of density
fluctuations. Experimental complications, including spontaneous emission
and unwanted ``rogue'' photodissociation from a photoassociated molecule are
analyzed. A two-color Raman scheme is studied as a method to set up an
effective two-mode scheme with reduced spontaneous emission losses. We
discuss photoassociation rates and photoassociation Rabi
frequencies for high-lying vibrational states in alkalis both on the basis
of molecular-structure calculations, and by comparing with an experiment
[Wynar {\it et al}., Science {\bf 287}, 1016 (2000)].
\end{abstract}

\maketitle

\section{Introduction}

Laser cooling and its spin-offs now routinely produce gaseous samples in
which thermal energies, when expressed as frequencies, are smaller than the
typical linewidth of an optical dipole transition. Photoassociating
transitions, in which two thermal atoms combine in the presence of light to
make a molecule, may therefore exhibit linewidths every bit as narrow as
the transitions one encounters in nonlinear laser spectroscopy. As a
result, photoassociation spectroscopy~\cite{PATHEO} has become the source
of the most accurate molecular structure data available. Bose-Einstein
condensation~\cite{BEC} is another recent triumph in the quest toward low
temperatures in atomic physics. The connection between photoassociation and
Bose-Einstein condensation has long been close, though somewhat incidental;
photoassociation spectroscopy has provided  key numerical data for
condensation experiments~\cite{SCALEN}.

There have been early discussions of photoassociation of a condensate
itself~\cite{BUR97,JUL98}. However, to us, the true scope of the
connection between photoassociation and condensation was only revealed by
our explicit observation~\cite{JAV98} that in a thermal gas it is the
same phase space density that governs both the onset of Bose-Einstein
condensation {\em and\/} the efficiency of photoassociation.

At the heart of our photoassociation work lies the quasicontinuum (QC)
approach~\cite{JAV98,MAC99}. The idea is to enclose two colliding atoms in
a box, which has the effect of discretizing the dissociation continuum of
the corresponding diatomic molecule. At the end of the calculations,
the quantization volume is taken to infinity. Aside from resolving
certain mathematical difficulties, this method turns out to have the
unexpected benefit that analysis of photoassociation is reverted to
analysis of few-level systems, as in quantum optics or laser spectroscopy.
For instance, studies of two-color photoassociation schemes may draw from
decades of experience in quantum optics and laser spectroscopy~\cite{MAC99}.

However, in its initial form our QC approach does not apply to a quantum
degenerate sample. We sought to rectify this shortcoming by introducing a
phenomenological second-quantized Hamiltonian for
photoassociation~\cite{JAV99}. This idea was developed at the same time
independently by Drummond {\it et al}.~\cite{DRU98}, and mathematically
closely related approaches to the Feshbach resonance are also under active
study~\cite{TIM98,TIM99,ABE99}. Comparison with the QC approach gives the
transition matrix elements to insert into our Hamiltonian.

We first considered a two-mode model that only takes into account one C.M.
wave function for atoms and one for molecules, the modes containing the
atomic and molecular condensates~\cite{JAV99}. The main finding was coherent
photoassociation analogous to coherent transients in few-level systems. For
instance, the system may exhibit a form of Rabi flopping between atoms and
molecules. Moreover, by properly sweeping the frequency of the
photoassociating laser, in a process akin to rapid adiabatic passage the
atomic condensate may be turned into a molecular condensate~\cite{JAV99}.
In another development in this direction, we have argued that two-color
free-bound-bound stimulated Raman adiabatic passage, \hbox{STIRAP}, is
feasible starting from an atomic condensate~\cite{MAC00}. We have also
gone beyond the two- and three-mode approximations, allowing for an
arbitrary position dependence of the atomic and molecular condensates,
albeit in a classical approximation similar to the one underlying the
Gross-Pitaevskii equation~\cite{JAV99a}. It then turns out that an
equilibrium with both atomic and molecular condensates present together
with the photoassociating light is unstable. The sample tends to collapse
spontaneously into clumps whose densities increase with
time~\cite{JAV99a}.

The primary purposes of the present paper are to document the numerous
technical and physical details of our second-quantized approach to
photoassociation that could not be accommodated by the letter format of
Refs.~\cite{JAV99}, \cite{MAC00}, and~\cite{JAV99a}, and to extend our
discussion in several directions that support those references. To offer a
comprehensive account of the field theory version of our approach to
coherent photoassociation, we have found it necessary to analyze the dipole
matrix element for photoassociation in detail. This endeavor in effect
constitutes an alternative derivation (c.f. Ref.~\cite{MAC99}) of our
entire QC methodology. Second, we add an analysis of two-color
photoassociation of a quantum-degenerate sample in a three-mode
approximation that is to some extent complementary to the one in
Ref.~\cite{MAC00}. Much as expected, the two-color scheme provides a
reprieve from spontaneous-emission losses from the primary photoassociated
state.  Third, we present a quantitative analysis of ``rogue''
photodissociation from a molecular condensate to atomic modes outside the
condensate. Our suggestion~\cite{JAV99} that with increasing light
intensity the unwanted photodissociation may overtake coherent
condensate-condensate transitions  is corroborated. We find a minimum
usable time scale proportional to the the inverse of the recoil frequency
of laser cooling.

Probably the most prominent qualitative finding emerging from our work is
the observation that it is Bose enhancement that ultimately facilitates
coherent transients such as Rabi flopping, adiabatic following, and STIRAP
in photoassociation of a condensate~\cite{MAC00,MAC00a}. Throughout this
paper, we continue to demonstrate how coherent optical transients come
about in a condensate, and argue why they should be absent in a
nondegenerate gas.

There has recently been a remarkable experiment on two-color
photoassociation of a condensate~\cite{WYN00}. Accordingly, we include
a detailed discussions on the values of experimental parameters in alkalis
in general, and the parameters of Ref.~~\cite{WYN00} in particular. The
analysis of the actual experiment demonstrates that there is still some
way to go before genuinely coherent photoassociation is reached.

In Sec.~\ref{FTA} we give a walk-through of our second-quantized
Hamiltonian, including a detailed discussion of the dipole moment matrix
element and both the momentum and position representation of the
Hamiltonian. The special case with only one spatial mode for both atoms and
molecules is the subject of Sec.~\ref{2MODEMODEL}. The classical
version of the field theory, but including all spatial modes, is the
subject of Sec.~\ref{FTALL}. The numerous complications to our one color
scheme that one is liable to encounter in real experiments, as well as
the experimental parameter values, are the subject of Sec.~\ref{EXCON}. The
brief remarks in Sec.~\ref{CONCL} conclude the paper. There are also two
appendices, \ref{DMA} on the details of the relation of the dipole matrix
element between second-quantized and quasicontinuum approaches, and
\ref{PAK} on the role of atom-atom collisions in our development.

\section{Field theory for atoms and molecules}\label{FTA}

The task of the present section is to develop in detail the
second-quantized approach governing photoassociation of atoms into 
molecules in the
prototype case of one laser color only. Simple heuristic arguments based on
Refs.~\cite{JAV98,MAC99} could, and did, achieve most of our aims in
Ref.~\cite{JAV99} where we dealt primarily with the momentum
representation, but the field theory of Ref.~\cite{JAV99a} calls for a few
additional angles. To support them, we present here a partially new
ab-initio discussion of our QC method.

We take the photoassociating atoms to be in precisely one internal state,
and similarly we assume that photoassociation leads to molecules with 
precisely one
internal state. These assumptions could be relaxed, but then one has to
follow the fate of the internal states as well. We do not go into
this, but in essence assume that (i) the atoms are polarized and that
(ii) the photoassociation resonance in itself selects a unique final 
state for the
molecule.

\subsection{Two atoms}\label{2ATOM}

We begin with a pair of atoms, assumedly in the dissociation continuum of
a given potential energy curve of a diatomic molecule. As the atoms
interact, their relative momentum need not be a constant of the motion.
Nonetheless, given a finite range for atom-atom interactions, in
free space the wave functions of the relative motion $\phi_\kvec(\rvec)$ could
still be characterized by the asymptotic ($r\rightarrow\infty$) wave vector
$\kvec$.

On the other hand, in the spirit of the QC method~\cite{JAV98,MAC99},
we assume that the relative motion of the atoms is confined to a finite
volume $V$. There are two basic questions in our two-atom analysis that
must be considered.  First, our phenomenological many-particle Hamiltonian
(Ref.~\cite{JAV99} and Eq.~\eq{HK} below) is written down in terms of
plane waves, yet it is more common to analyze photoassociation in terms of
angular-momentum partial waves. How do we make the connection? Second,
in our quasicontinuum method we resort to a finite quantization volume
$V$, which tends to infinity only at the end of the calculations. How
should we handle the finite quantization volume?

We quantize the relative motion of the two atoms in a spherical box of
radius $R$ and volume $V={4\over3}\pi R^3$ using reflecting boundary
conditions. While angular momentum is still a constant of the motion,
the usual eigenstates~\cite{MAC99} of the relative motion cannot be
characterized by a momentum vector, even asymptotically.  Nonetheless, it
is evidently possible to construct orthonormal superpositions
$\bar\phi_\kvec(\rvec)$ of the eigenstates of the spherical box that in the
limit of a large box turn into the states $\phi_\kvec(\rvec)$, i.e., states
that behave like plane waves at large distances.

Let us first take the inner product of a true plane wave, normalized to
the volume $V$, with a spherically symmetric (real) test function $f(r)$
with a finite range $\ll R$. In the limit $\kvec\rightarrow0$ we have
\beq
I_1 = \lim_{\kvec\rightarrow0}\,{1\over\sqrt{V}} \int_0^R
d^3r\,e^{i\kvec\cdot\rvec}f(r) \simeq {1\over\sqrt{V}}
\int_0^\infty 4\pi r^2\,dr\,f(r)\,.
\eeq
Next we take the same inner product for the $l=0$ partial wave of the
plane wave $e^{i\kvec\cdot\rvec}$, basically the spherical Bessel function
$j_0(kr)\propto \sin kr /r$. We normalize this partial wave also in the
spherical box, or equivalently, in the radial coordinate
$r$ with respect to the measure $4\pi r^2\,dr$. In the limit of small $k$
we have the integral for the inner product
\bea
I_2 &=& {1\over\sqrt{2\pi R}}\int_0^R 4\pi r^2\,dr\,{\sin
kr\over r}f(r)\nonumber\\
&\simeq&{k\over \sqrt{2\pi R}}\int_0^\infty4\pi
r^2\,dr\,f(r)\,.
\eea
Obviously, the ratio of the two,
\beq
\alpha(k) = {I_1\over I_2} = {1\over k}{\sqrt{ 2\pi R\over V}}\,,
\label{CF}
\eeq
is the expansion coefficient of the $s$-wave in the plane wave, given
that both the plane wave and the $s$-wave are normalized to the volume
$V$. Actually, the $k$ values of the spherical eigenmodes are quantized,
and the smallest value is  $k=\pi/R$. We therefore have to be careful with
the limit $k\rightarrow0$ when applying Eq.~\eq{CF}.

When atom-atom interactions are taken into account, the radial eigenstates
for a given angular momentum are not just spherical Bessel functions, but
also reflect the scattering phase shifts. Henceforth we assume that
only $s$-wave scattering needs to be considered, as is usually the case for
bosons at sufficiently low temperatures. Obviously, even though the partial
waves have phase shifts, by tweaking the phases of the partial waves it is
still possible to make an asymptotic near-plane wave $\bar\phi_k(\rvec)$ out
of the eigenstates of the relative motion of the atoms in the sphere. The
adjustment of phases has no effect on the weight of the $s$-wave component
in $\bar\phi_k(\rvec)$, which may still be inferred from Eq.~\eq{CF}.

\commentout{
Now, the total density of states in spherical box for a given wave number
of relative motion of the atoms $k$ is the same as in a cubic box, namely,
\beq
D(k) ={k^2V\over 2\pi^2}\,,
\eeq
and for a given $k$ these are spread out over
\beq
n(k) = \int_0^{kR}dl\,(2l+1) \simeq (kR)^2
\eeq
angular momentum eigenstates. Here $V={4\over3}\pi R^3$ is the
quantization volume. As we must not gain or lose degrees of freedom, the
would-be asymptotic momentum eigenstates $\bar\phi_\kvec(\rvec)$ have to obey
the same counting.

The difference is that, rather than being angular-momentum eigenstates,
the  states $\bar\phi_\kvec(\rvec)$ are superpositions of angular-momentum
eigenstates. For further guidance, let us consider the standard expansion
of a plane wave propagating in the $z$ direction in terms of
angular-momentum eigenstates,
\bea
e^{ikz}&=&\sum_{l=0}^\infty(2l+1)i^lj_l(kr)P_l(cos\theta)\nonumber\\
&=&\sum_{l=0}^\infty\sqrt{2l+1\over 4\pi}j_l(kr)Y_l^0(\theta,\varphi)\,.
\eea
We have written this in terms of the spherical harmonics, the orthonormal
function on the surface of a sphere. For our considerations of the states
$\bar\phi_\kvec(\rvec)$, two aspects are different. First, the sum obviously
must cut off at $l=kR$, not run off to infinity.  Second, the wave vector
$\kvec$ does not have to point in the $z$ direction. We must then perform a
coordinate rotation that converts each $Y_l^0$ into a normalized
superposition of the functions $Y_l^m$. The relevant message, though, is
the following: No matter what the direction, the angular momentum $l$
appears in the sum with the weight $2l+1$. Thus, given the maximum angular
momentum $kR$, the weight of the $l=0$ component in $\bar\phi_k(\rvec)$ is
always $1/n(k)=1/(kR)^2$.}

Let us denote the wave function of the particular molecular state we are
aiming for by $\bar\psi(\rvec)$, and make the standard (albeit crude)
approximation that the relevant electronic dipole matrix element $\db$ is a
constant independent of the relative coordinate $\rvec$ of the atoms
comprising the molecule. Then the QC dipole matrix element for
photoassociation and photodissociation characterizing a process in which
the relative momentum of the colliding atoms is
$\hbar\kvec$ reads simply
\beq
\db(\kvec) = \db\int d^3r\,\bar\psi^*(\rvec)\bar\phi_\kvec(\rvec) \,.
\label{MEL}
\eeq
We only consider $s$-wave collisions, and correspondingly set
$\bar\psi(\rvec)=\bar\psi(r)$ as appropriate for a nonrotating
$J=0$ molecule. Of course, only the $l=0$ component of the wave function
$\bar\phi_\kvec(\rvec)$ counts. Taking into account the weight from
Eq.~\eq{CF}, we have
\beq
\db(\kvec) = \db\,{1\over k}{\sqrt{ 2\pi R\over V}}
\int_0^R 4\pi
  r^2\,dr\,\bar\psi^*(r)\bar\phi_k^0(r)\,.
\label{D1}
\eeq
The notation $\bar\phi^0_k(r)$ stands for the radial $l=0$ wave function
of the relative motion of the two atoms corresponding to the wave number
$k$, and normalized to volume $V$ as usual.

A convenient qualitative model is provided by the limiting form
\beq
\bar\phi^0_k(r) = {1\over\sqrt{2\pi R}}\,{\sin\,k(r-a)\over r}\,,
\label{WFQ}
\eeq
where $a$ is the $s$-wave scattering length. Combination of \eq{D1}
and~\eq{WFQ} gives
\beq
\db(\kvec) = \db\,{4\pi\over k\sqrt{V}} \int_0^R
r\,dr\,\sin\,k(r-a)\bar\psi^*(r)\,.
\label{D2}
\eeq
The form~\eq{WFQ} is only valid outside the range of the atom-atom
interaction potential, and for small enough $k$ so that the scattering
phase shift may be written as $ka$. Equation~\eq{D2} is therefore
quantitatively reliable only if the vibrational bound-state wave function of
the molecule $\bar\psi(r)$ happens to reside outside the range of the
atom-atom interactions of the photoassociating atoms. In fact, this is
the case in some of the current photoassociation
experiments~\cite{COT95,COT95a}.

The matrix element~\eq{D2} is explicitly proportional to $V^{-1/2}$.
In fact, the $V^{-1/2}$ scaling of $\db(\kvec)$ is a generic
property of our QC approach, and does not depend on the specific assumptions
of Eq.~\eq{D2}. First, the almost-plane wave
$\bar\phi_\kvec(\rvec)$ is normalized to the volume
$V$. No matter what kind of (square integrable) structure occurs  around
$\rvec\sim0$, the plane-wave form at large distances makes the normalization
constant proportional to $V^{-1/2}$. Second, because of the finite range of
the bounded molecular wave function $\bar\psi(r)$, its normalization
coefficient does not depend on $V$. Third, thanks to the same finite
range, the integral~\eq{MEL} effectively extends only over a finite, fixed
volume. The net result is that the matrix element indeed is proportional to
the normalization constant of $\bar\phi_\kvec(\rvec)$, $\propto V^{-1/2}$.

Another important property of the matrix element~\eq{D2}, that it
tends to a constant as $k\rightarrow0$, is also a generic feature
of the physics. Namely, for small enough $k$, the form of the $s$-wave
$\bar\phi^0_k(r)$ indeed is $\sin k(r-a)/r$, except within the range of
atom-atom interactions. But within this range, the shape of
$\bar\phi^0_k(r)$ is independent of $k$ at small enough $k$. In order that
the inner and the outer forms join smoothly, the amplitude of the inner wave
function must be $\propto k$ for small $k$, just as is the amplitude of the
outer wave function. So, for small enough
$k$, the wave function
$\bar\phi^0_k(r)$ is $\propto k$ for all of those $r$ for which the
bound-state wave function $\bar\psi(r)$ is effectively nonzero. The matrix
elements~\eq{MEL}, \eq{D1} and~\eq{D2} therefore all tend to a
constant as $k\rightarrow0$.

In order to prepare for the comparison between photoassociation and
photodissociation, let us assume that a laser field with the amplitude
\beq
\E(t) = \half\E e^{-i\omega t}  + {\rm c.c.}
\eeq
is incident on a molecule. We denote the detuning of the light above the
photodissociation threshold by $\delta$. Given the reduced mass of the
colliding atoms $\mu$, the corresponding resonant wave number $k_0$ and
velocity $v_0$ are such that $\hbar k_0^2/2\mu = \delta$ and
$v_0=\hbar k_0/\mu$. Using the standard dipole and rotating-wave
approximations as well as Fermi's golden rule, we have the 
photodissociation rate
\bea
\Gamma_0 &=& {2\pi\over\hbar^2}\int dk\,D(k)
\left|{\db(\kvec)\cdot\E\over 2} \right|^2\delta\left({\hbar
k^2\over2\mu}-\delta\right)\nonumber\\
&=&{\mu^2 v_0 V\over\pi\hbar^2}\left|{\db(k_0)\cdot\E\over
2\hbar}
\right|^2\,,
\label{G0}
\eea
where
\beq
D(k) ={k^2V\over 2\pi^2}\,
\eeq
is the density of $k$ states, whether in a cubic or in a spherical
box~\cite{MAC99,VOLREF}.
Equation~\eq{G0} is fully compatible with the development
in Refs.~\cite{JAV98} and~\cite{MAC99}, as it should.

Since the dipole matrix element tends to a constant in the limit
$k\rightarrow0$, it is easy to see from~\eq{G0} that the Wigner
threshold law holds; namely, that $\lim_{v_0\rightarrow0}\Gamma_0/v_0$ is
finite, and nonzero except for an unlikely accident. As a matter of
fact, our argument about the $k\rightarrow0$ limit of the dipole matrix
element was nothing but a recital of a standard argument for the
Wigner threshold law. Nonetheless, it will furnish a relevant piece
of the puzzle when we are to discuss atom-molecule field theory below. To
this end, we note from Eqs.~\eq{D1} and~\eq{G0} the equality

\bea
\lefteqn{\lim_{v_0\rightarrow0} {\Gamma_0\over v_0}}\nonumber\\
&=&
{\mu^2\over\pi\hbar^2}
\left|
{\db\cdot\E\over\hbar}\,\lim_{k_0\rightarrow0}{\sqrt{2\pi R}\over
k_0}\,\int 4\pi r^2\,dr\,\bar\psi^*(r)\bar\phi^0_{k_0}(r)
\right|^2\,.\nonumber\\
&{}&
\label{GC}
\eea

\subsection{Many atoms in momentum representation}
\subsubsection{Basic Hamiltonian}

In Ref.~\cite{JAV99} we introduced a phenomenological second-quantized
Hamiltonian for photoassociation of bosonic atoms to (obviously) bosonic
molecules,
\bea
{H\over\hbar} &=& \sum_\kvec\left[ {\hbar\kvec^2\over4m}b^\dagger_\kvec
b_\kvec +
\left(
-{\delta_0\over2}+{\hbar \kvec^2\over 2m}\right) a^\dagger_\kvec a_\kvec
\right]\nonumber\\
&&-\sum_{\kvec\kvec'\q}\left[{\db_{\kvec\kvec'}
\cdot\E_\q\over4\hbar}b^\dagger_{\kvec+\ 
k'+\q}a_\kvec
a_{\kvec'} + {\rm H.c.}\right]\,.
\label{HK}
\eea
Here $m$ stands for the mass of the atom, $m=2\mu$. The operators $a_\kvec$
and $b_\kvec$ are boson annihilation operators for atoms and molecules in the
plane wave mode $\kvec$. The Hamiltonian trivially includes the kinetic energy
of the atoms and molecules. Here photoassociation takes place with an
absorption (as opposed to induced emission) of a photon. By momentum
conservation, atoms with wave vectors $\kvec$ and $\kvec'$ plus a photon with wave
vector $\q$ must then make a molecule with wave vector $\kvec+\kvec'+\q$. This
explains the form of the cubic operator product; annihilate the atoms and a
photon, create the molecule.

As is usual in quantum optics, we use a classical field to represent the
photons. Specifically, the positive-frequency part of the electric
field reads
\beq
\E^+(\rvec) = \half \sum_\q \E_\q\,e^{i\q\cdot\rvec}\,.
\label{EPLUS}
\eeq
Both $\E^+(\rvec)$ and the coefficients $\E_\q$ may in principle be slowly
varying functions of time, but the leading time dependence of the
electric field $\propto e^{-i\omega t}$ has been absorbed into the
detuning $\delta_0$ in a  transformation to a rotating frame.
The seemingly unexpected factor $\half$ in the detuning term correspond to
the fact that upon photodissociation one molecule produces two atoms, both
of which generally take away kinetic energy with them. The sign of the
detuning is chosen in such a way that
$\delta_0>0$ corresponds to tuning of the laser by the energy
$\hbar\delta_0$ above the photodissociation threshold. A quick way to 
verify this is to
consider the potential resonance when the (quasi) energy (in the rotating
frame) for a system of one molecule and zero atoms would be the same as the
energy for a system with zero molecules and two atoms, much like in the
Appendix~\ref{DMA}. The total energy of the atoms must equal the energy of
the molecule plus $\hbar\delta_0$, which is compatible with Eq.~\eq{HK}.

We assume that the optical transition responsible for photoassociation is a
dipole transition. In the dipole approximation, the electronic 
photodissociation and photoassociation
transitions in a molecule do not (cannot!) depend on the propagation
direction of light, hence the dipole matrix elements $\db_{\kvec\kvec'}$
do not depend on photon momenta. Besides, by translational invariance, the
dipole matrix elements must be functions of the difference $\kvec-\kvec'$
only. Because of the exchange symmetry of the boson operators, the matrix
element
$\db_{\kvec\kvec'}$ may be chosen to be symmetric in the exchange of the
momentum indices, hence, an even function of $\kvec-\kvec'$. And finally,
since we consider $s$-wave photoassociation only, the matrix element is a
function of $|\kvec-\kvec'|$.

To pin down the matrix elements $\db_{\kvec\kvec'}$, in Ref.~\cite{JAV99}
we took the limit of a dilute thermal gas. The known QC results are
recovered if one expresses the thus far undefined coefficients
$\db_{\kvec\kvec'}$ in terms of the matrix elements~\eq{MEL} as
\beq
\db_{\kvec\kvec'}\equiv\sqrt{2}\,\db\left(\hbox{$1\over2$}(\kvec-\kvec')\right)\,.
\label{ID}
\eeq
The $\sqrt{2}$ is a consequence of the Bose-Einstein statistics, and the
factor $1\over2$ follows from the way that the relative momentum must be
defined to make it the conjugate of the conventional relative position.
We have recently noticed that there is a subtlety associated with this
identification having to do with the the statistics of the atoms. We
elaborate in Appendix~\ref{DMA}, but meanwhile continue according to
Eq.~\eq{ID}.

In sum, if one may treat the atoms and the molecules as bosons in their
own right, then both the form of, and even the numerical coefficients in,
the photoassociation Hamiltonian are unambiguously determined by simple
physical considerations. Admittedly, we do not know of any conclusive
{\it ab initio\/} argument for the bosonic nature of atoms and molecules
in  photoassociation.
But neither do we know of any for alkali atoms, which nonetheless seem
to behave like good bosons in current BEC experiments.

\subsubsection{Two-mode approximation}

We now revisit the situation that was the focus of Ref.~\cite{JAV99}: an
infinite, homogeneous condensate and one plane wave of light with photon
momentum $\hbar\q$. By conservation of momentum, the  molecules one may
create by photoassociation all have the wave vector equal to~$\q$. On the
other hand, as far as momentum conservation is concerned, a molecule with
the wave vector $\q$ may photodissociate into two atoms, neither of which
has zero momentum. We will discuss such ``rogue photodissociation'' in more
detail below, Sec.~\ref{ROGUE}, and so far ignore it.

All told, we only retain the two modes of the model with the annihilation
operators $a\equiv a_0$ and $b\equiv b_\q$. The Hamiltonian reads
\bea
{H\over\hbar} &=& {\hbar\q^2\over4m}\,b^\dagger b -\half\delta_0\,
a^\dagger a
  - \half\kappa (b^\dagger a a + b a^\dagger a^\dagger)\nonumber\\
&\simeq&{\hbar\q^2\over4m}\,b^\dagger b -\half\delta_0\,a^\dagger
a - \half\kappa (b^\dagger a a + b a^\dagger a^\dagger) \nonumber\\
&&-
{\hbar\q^2\over8m}(2 b^\dagger b + a^\dagger a)\nonumber\\
&\equiv&-\half\delta\,a^\dagger a
  - \half\kappa (b^\dagger a a + b a^\dagger a^\dagger)\,.
\label{2MH}
\eea
In the ``approximate'' equality we have added a constant of the motion to
the Hamiltonian, a step that has no effect on the ensuing dynamics. The
QC Rabi frequency reads
\bea
\kappa &=& {\sqrt{2}\over
2\hbar}\,\E\cdot[\lim_{\kvec\rightarrow0}\db(\kvec)]\nonumber\\
&=& \lim_{v_0\rightarrow0} \sqrt{2\pi\hbar^2\Gamma_0\over\mu^2 V v_0}\,,
\label{RAB}
\eea
where we have used Eq.~\eq{G0}. Without any loss of generality, we have
chosen $\kappa$ to be real and nonnegative. Seemingly alarmingly, the Rabi
frequency still depends on the quantization volume. We will return to this
point in Sec.~\ref{2MODEMODEL}. At present, our aim is just to set up the
second-quantized Hamiltonian in the two-mode approximation. The task is
completed by noting that
\beq
\delta = \delta_0+{\hbar\q^2\over4m}
\label{DDEF}
\eeq
is the detuning corrected for the photon recoil energy of the molecule.
 From now on we will keep track of this distinction, so that $\delta$ always
includes the appropriate recoil. In fact, that was already implicitly the
case in Eq.~\eq{G0}.

\subsection{Many-atom field theory}

Our phenomenological Hamiltonian~\eq{HK} was written down originally in
momentum representation. Nevertheless, as in Refs.~\cite{JAV99}
and~\cite{JAV99a}, it is fairly straightforward to convert it into
position representation, i.e., into a quantum field theory.

Let us introduce atomic and molecular fields in a quantization volume
$V$. {\it A priori}, this volume does not have to be the cavity of radius
$R$, as in Sec.~\ref{2ATOM}. Often it is actually more convenient to use a
cubic box with periodic boundary conditions. On the other hand, when one
deals with two atomic fields, it is often expedient to write the integrals
in the theory in terms of center-of-mass and relative coordinates. If the
integral over the relative coordinates cuts off because the integrand tends
to zero at large distances, it is immaterial what volume is used, as long
as it is large enough. Thus,  when convenient, in such relative-coordinate
integrals we may still imagine the spherical potential well. The short of
the story is that the quantization volume $V$ refers to the geometry
that is expedient in the particular context.

We write the atomic and molecular fields in terms of plane wave states as
\beq
\phi(\rvec) = {1\over\sqrt{V}}\sum_\kvec e^{i\kvec\cdot\rvec} a_k,\qquad
\psi(\rvec) = {1\over\sqrt{V}}\sum_\kvec e^{i\kvec\cdot\rvec} b_k\,.
\label{TOFIELDS}
\eeq
The Hamiltonian~\eq{HK} may be cast as a Hamiltonian density of
a field theory for these fields. Using the standard continuum limit
\beq
\sum_\kvec f(\kvec)\simeq {V\over(2\pi)^3}\int d^3k \, f(\kvec)\,,
\eeq
Eq.~\eq{EPLUS}, and the properties of Fourier integrals,  the part
of the Hamiltonian density depending on the dipole interaction becomes
\beq
{{\cal
H}(\rvec)\over\hbar}=\!-\!{\E^+(\rvec)\over2\hbar}\cdot\psi^\dagger(\rvec)\!\int
\!\!d^3r'\,\phi(\rvec\!+\!\half\rvec')\db(\rvec')\phi(\rvec\!-\!\half\rvec')
+
\ldots.
\eeq
The dipole kernel of photoassociation $\db(\rvec)$ is given in terms of the
dipole matrix element of Eq.~\eq{MEL} as
\beq
\db(\rvec) = {\sqrt{2V}\over (2\pi)^3}\int
d^3k\,e^{-i\kvec\cdot\rvec}\,\db(\kvec)\,.
\label{DR}
\eeq
It should be noted that our argument contains a subtle trick, which is
exposed in Appendix~\ref{PAK}.

To make further progress, we study the kernel using the qualitative
model for the dipole matrix element~\eq{D2}. Some more juggling with
Fourier transforms gives

\bea
\lefteqn{\db(\rvec) }\nonumber\\
&=& \sqrt{2}\,\db
{
\theta(a\!+\!r)(a\!+\!r)\bar\psi^*(a\!+\!r)\!-\!\theta(a\!-\!r)(a\!-\!r)
\bar\psi^*(\!a-\!r)\over
r }\,,\nonumber\\
\label{DR1}
\eea
where $\theta$ is the Heaviside unit step function.
This result is only as good as the assumptions of Eq.~\eq{D2}. In
particular, as Eq.~\eq{D2} is not valid for large $k$, Eq.~\eq{DR1} is
dubious at short distances. Also, as we already pointed out
with  Eq.~\eq{D2},  even for small
$k$ the model is quantitatively accurate only if the vibrational wave
function $\bar\psi(r)$ resides outside the range of the atom-atom
interactions. In spite of these caveats, Eq.~\eq{DR1} gives an
exceedingly plausible idea of the {\em range\/}
$\Delta  r$ of the dipole interaction kernel; $\Delta  r$ is of the order of
the larger of the spatial extent of the vibrational state $\bar\psi(r)$ and
the absolute value of the scattering length $a$.

The relevant length scale for the atomic field is larger than
$\Delta r$ if the energies of all of the atoms relevant for the time
evolution are smaller than $\sim\hbar^2/m(\Delta r)^2$. Henceforth we
assume this to be the case. Then we may simplify the field-theory form of
the integral as follows,
\begin{widetext}
\bea
{{\cal
H}(\rvec)\over\hbar}&=&-{\E^+(\rvec)\over2\hbar}\cdot\psi^\dagger(\rvec)\int
d^3r'\,\phi(\rvec+\half\rvec')\db(\rvec')\phi(\rvec-\half\rvec') +
\ldots\label{KERFOR}
\simeq-{\E^+(\rvec)\over2\hbar}\cdot\psi^\dagger(\rvec)\,\phi(\rvec)\phi(\rvec)
\int d^3r'\db(\rvec') + \ldots\nonumber\\
&=& -\left[{\E^+(\rvec)\over2\hbar}\cdot
\sqrt{2V}\,\lim_{\kvec\rightarrow0}\db(\kvec)
\right]
\psi^\dagger(\rvec)\,\phi(\rvec)\phi(\rvec)+\ldots\nonumber\\
&=&
\left[\sqrt{2}\,{\db\cdot\E^+(\rvec)\over2\hbar}
\lim_{k\rightarrow0}{\sqrt{ 2\pi R}\over k}
\int 4\pi
  r^2\,dr\,\bar\psi^*(r)\bar\phi_k^0(r)
\right]
\psi^\dagger(\rvec)\,\phi(\rvec)\phi(\rvec)+\ldots\,.
\eea
\end{widetext}
The second equality follows from the assumedly slow dependence of the
atomic field $\phi(\rvec)$ on position $\rvec$ compared to $\db(\rvec)$, the
third from Eq.~\eq{DR} and the properties of Fourier transformations, and in
the last equality we have substituted Eq.~\eq{D1}. Comparison with
Eq.~\eq{GC} then gives the contact interaction form of the photoassociation
dipole interaction,
\beq
{{\cal H}(\rvec)\over\hbar} = - {\cal D}(\rvec)
\psi^\dagger(\rvec)\,\phi(\rvec)\phi(\rvec)+\ldots\,,
\label{CIF}
\eeq
with
\beq
{\cal D}(\rvec) = e^{i\,\arg[\db\cdot\E^+(\rvec)]}\,\lim_{v_0\rightarrow0}
\sqrt{2\pi\hbar^2\Gamma_0(\rvec)\over v_0 \mu^2 }\,.
\eeq
Here $\Gamma_0(\rvec)$ is the photodissociation rate as per the prevailing
light intensity at position $\rvec$, and the exponential records the
phase of the dipole interaction term.

For reference, we summarize the contact-interaction version of the
Hamiltonian for photoassociation in terms of atomic and molecular 
fields. While at that,
by adding a suitable multiple of the conserved particle number completely
analogously to the operation that we did in the chain of
equations~\eq{2MH}, we move the detuning term to the molecules. The result
is
\bea
H&=&\int d^3r\,{\cal H}(\rvec);\nonumber\\
{{\cal H}(\rvec)\over\hbar} &=&
\phi^\dagger(\rvec)\left(-{\hbar\nabla^2\over2m}\right)\phi(\rvec)
+\psi^\dagger(\rvec)\left(-{\hbar\nabla^2\over4m}+\delta_0\right)
\!\psi(\rvec) 
\nonumber\\
&&- \left[ {\cal D}(\rvec)  \psi^\dagger(\rvec)\,\phi(\rvec)\phi(\rvec) + {\rm
H.c.}\right]\,.
\label{HF}
\eea

\section{Coherence in two-mode model}\label{2MODEMODEL}

We consider the two-mode model, whose Hamiltonian we reproduce here from
Eq.~\eq{2MH},
\[
{H\over\hbar} =  -\half\delta\,a^\dagger a
  - \half\kappa (b^\dagger a a + b a^\dagger a^\dagger)\,.
\]
This is the age-old Hamiltonian for second-harmonic generation,
with atoms replacing the fundamental-frequency light and molecules
the second harmonic. Needless to say, the literature on the model is
extensive. We will cite a few relevant examples below.

Let us take the conserved particle number  $a^\dagger a + 2b^\dagger b$
to have the value $N$. The Hamiltonian~\eq{2MH} may then be restricted to
the space spanned by the states $|n\rangle \equiv |n\rangle_M|N-2n\rangle_A$
  with $n =0,\ldots, N/2$ molecules and $N-2n$ atoms, and
is tridiagonal in that basis. It is easy to find both the eigenstates of the
Hamiltonian and the time evolution of any specified initial state
numerically. We have done so using inverse iteration, and a variation of
the Crank-Nicholson method. While the problem considered in
Ref.~\cite{JAV99b} is different, the numerical methods described therein
work equally well here.

In this way we have firstly~\cite{JAV99,WAL72} found the fraction of atoms
converted into molecules, given that the system starts out at time $t=0$
with everything in atoms and is driven by a resonant laser, $\delta=0$. We
found a nonlinear analog of Rabi flopping, the system oscillating between
atoms and molecules. Given that we have a quantum system where the spacing
between the successive energy eigenstates is not constant, it is not a
surprise that the oscillations collapse, and even revive~\cite{DRO92}.

Interesting as these features are as a matter of principle, the drawback
remains that Rabi oscillations of occupation probabilities are
generally not robust, not even in quantum optics or laser spectroscopy. We
foresee little experimental utility for nonlinear Rabi oscillations. Our
main message of the analysis of Rabi oscillations in Ref.~\cite{JAV99}
rather is that, for $N\gg1$, the characteristic frequency scale of the
system is not the QC Rabi frequency $\kappa$ but
\beq
\Omega = \sqrt{N}\,\kappa =
\lim_{v_0\rightarrow0} \sqrt{2N\pi\hbar^2\Gamma_0\over\mu^2 V v_0}
= \lim_{v_0\rightarrow0} \sqrt{2\pi\hbar^2\Gamma_0\rho\over\mu^2 v_0}\,.
\label{OME}
\eeq
The $\sqrt{N}$ is evidently a Bose enhancement factor. The implicit
quantization volume has disappeared from the result. Instead we have the
density of atoms if all molecules were dissociated,
\beq
\rho = {N\over V}\,,
\eeq
which is a true physical parameter for the system.

Bose enhancement highlights the role of statistics in our
analysis. Suppose we were to deal with a nondegenerate gas. Then the
occupation probability of each QC state is much less than unity by
definition, and there is no Bose enhancement. The frequency of Rabi
flopping between a given QC state and a  bound molecular states would just
be $\propto\kappa\propto V^{-1/2}$, and vanishes in the limit of large
volume. It is not possible to have Rabi flopping in photoassociation of an
infinite and homogeneous nondegenerate gas, even in principle~\cite{MAC00a}.

The photoassociation rate for a pair of atoms starting in a given QC state is
$\propto\kappa^2\propto 1/V$, and also vanishes in the limit of a large
volume. The saving grace~\cite{JAV98,MAC99} is that, to keep the density
constant, the atom number $N$ and at the same time the number of candidate
atoms to photoassociate with any given atom tends to infinity. Adding the
probabilities for photoassociation due to all colliders,  the total rate of
photoassociation for any given atom is $\propto 
N\kappa^2\propto\rho$. This remains a
constant in the continuum limit, which leads to a finite 
photoassociation rate $\propto
\rho$.

In contrast, in a BEC the atom number already ends up in the transition
matrix element, giving the effective Rabi frequency
$\Omega=\sqrt{N}\kappa\propto\sqrt{\rho}$. One does not add rates
$\propto\kappa^2$ due to different atoms, but all the condensate atoms act
as a single state that has a finite matrix element for 
photoassociation even in the
limit of infinite quantization volume. This ultimately facilitates
coherence effects like Rabi flopping.

\begin{figure}
\label{MOLFRA}
\includegraphics[scale=.77]{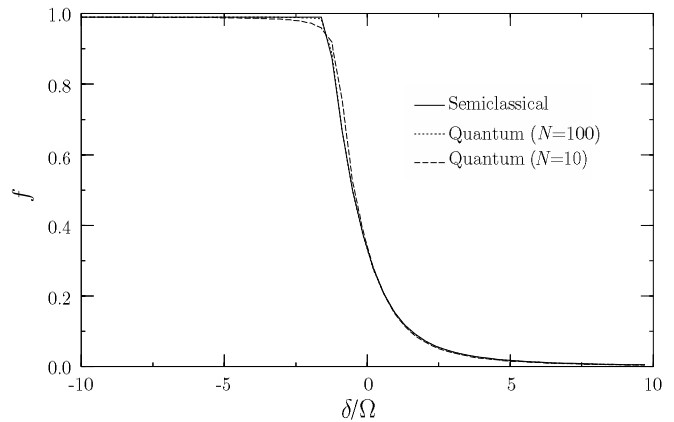}
\caption{
Fraction of atoms forming molecules, $f$, in the ground state of the
atom-molecule condensate as a function of the detuning of the laser from
photoassociation threshold, $\delta$. The long-dashed and short-dashed
lines are from the two-mode quantum treatment for the respective atom
numbers $N=10$ and $N=100$, the solid line is from the semiclassical
approach embodied in the Gross-Pitaevskii equations.}
\end{figure}

The Hamiltonian~\eq{2MH} acts in a rotating frame, and there is no
manifest physical significance to its eigenvalues. Nonetheless, let us
{\em call\/} them energies, and the state with the lowest energy the ground
state. Qualitatively, when the laser is tuned far below the
photodissociation threshold, $\delta\ll-\Omega$, the $a^\dagger a$ term in
the Hamiltonian becomes costly in energy. The ground state of the
Hamiltonian should then tend to have the system mostly in molecules.
Conversely, for $\delta\gg\Omega$, the ground state favors atoms. We
confirm these surmises explicitly in Fig.~\ref{MOLFRA} by plotting the
fraction of atoms converted into molecules, the expectation value
\beq
f={2\langle b^\dagger b\rangle\over N}\,,
\eeq
for the numerically obtained ground state. The difference between the
dashed curves is the atom number, $N=10$ and $N=100$.

The idea is near that, when the laser is swept from a
large above-threshold detuning to a large (in absolute value)
below-threshold value, the system will follow adiabatically, and an
atomic condensate is converted to a molecular condensates~\cite{JAV99}.
Moreover, if $\Omega$ were the relevant frequency scale, adiabaticity means
that the detuning must change by $\Omega$ in a time of the order or longer
than
$\Omega^{-1}$. In our model we assume that the detuning is swept as
\beq
\delta(t) = -\xi\Omega^2t\,.
\eeq
We expect adiabatic following when $\xi\lsimeq1$.

\begin{figure}
\includegraphics[scale=.60]{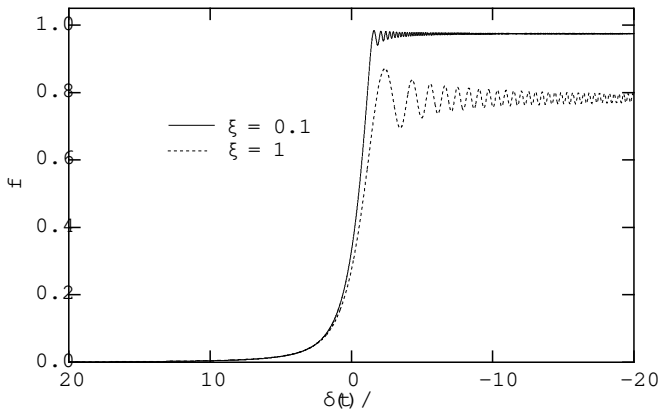}
\caption{Fraction of atoms converted to molecules,  $f=2\langle b^\dagger
b\rangle$, as a function of instantaneous detuning, $\delta(t)$, when the
detuning is swept linearly as a function of time according to
$\delta(t)=-\xi\Omega^2t$. The figure is for the atom number $N=100$, for
the two sweep rates $\xi=0.1$ and $\xi=1$.}
\label{SWEEP}
\end{figure}

This piece of intuition is correct. In Fig.~\ref{SWEEP} we reproduce
the relevant figure from Ref.~\cite{JAV99}. We fix $N=100$. At an
initial time giving $\delta = -20\,\Omega$, we start the system in its
ground state, almost purely atoms, and integrate the time dependent
Schr\"odinger equation numerically while sweeping the detuning at two rates
$\xi=1$ and $\xi=0.1$. The plot gives the fraction of atoms converted to
molecules as a function of the instantaneous detuning. For $\xi=1$ we expect
to be somewhere at the borderline of adiabaticity, and actually find a
conversion efficiency of 0.8. When the laser tuning is swept ten times more
slowly, $\xi=0.1$, the conversion efficiency has reached 0.97.

We believe that we have discovered a feasible method for preparing a
molecular condensate: start with an atomic condensate and a photoassociating
laser, then sweep the frequency of the laser in a proper way, and
obtain a molecular BEC. In the general manner of adiabatic methods, this
``rapid adiabatic passage'' should be robust.

Besides its potential utility, rapid adiabatic passage is another
example of coherent phenomena that occur in a condensate but not in a
nondegenerate gas. Once more, in a nondegenerate gas the frequency scale
for adiabaticity is $\kappa$, and the time scale is proportional to
$\kappa^{-1}\propto V^{1/2}$. Even if nothing else went wrong, in the limit
of a large sample the time scale for adiabatic atom-molecule conversion
would tend to infinity. Once more, Bose enhancement saves the day by
turning the volume dependence of the relevant frequency scale into a
density dependence~\cite{MAC00a}.

\section{Field theory for all modes}\label{FTALL}

If only one spatial mode has to be considered for both the atomic and the
molecular condensate, the two-mode Hamiltonian~\eq{2MH} is all there is to
it. Nonetheless, even if somehow the infinite homogeneous condensate could
be approximated in practice, for whatever reason there would always be atoms
and molecules present with momenta that are not included in the two-mode
picture. As the full Hamiltonian~\eq{HK} mixes momenta nonlinearly, the
possibility of instability arises and should be investigated.

We will study deviations from the two-mode picture, or the
possibility that the (infinite) atomic and molecular condensates are not
spatially homogeneous. Here we use the quantum field version of our 
photoassociation
theory. But solving nonlinear quantum field theories is generally not a
simple matter, and one must approximate. We will resort to classical field
theory, the same way one proceeds when an alkali condensate is analyzed
using the Gross-Pitaevskii equation.

\subsection{Gross-Pitaevskii equations}

To derive the Gross-Pitaevskii equations (GPE) for this problem, in the
usual way we first amend the Hamiltonian by adding to it a
multiple of the conserved particle number
\beq
N = \sum_\kvec(a^\dagger_\kvec a_\kvec + 2 b^\dagger_\kvec b_\kvec) = \int
d^3r\,[\phi^\dagger(\rvec)\phi(\rvec) + 2\psi^\dagger(\rvec)\psi(\rvec)]\,.
\eeq
Instead of the Hamiltonian density, we then use the ``Kamiltonian'' density
\beq
{\cal K}(\rvec) = {\cal H}(\rvec) - \mu[\phi^\dagger(\rvec)\phi(\rvec) +
2\psi^\dagger(\rvec)\psi(\rvec)]\,
\eeq
in our calculations.
This substitution has no effect on the dynamics. At this point the real
scalar $\mu$ is arbitrary. Though this piece of knowledge has no
bearing on our analysis, in the end the constant $\mu$ will be analogous to
chemical potential in thermodynamics. Hence, $\mu$ is called the chemical
potential.

Using the standard commutators for boson fields, the Heisenberg equation of
motion for the molecular field becomes
\bea
i\dot\psi(\rvec) &=& [\psi(\rvec),\int d^3 r'\,{{\cal
K(\rvec')}\over\hbar}\,]\nonumber\\
&=&\left[-{\hbar\nabla^2\over 4m} + \delta_0-2{\mu\over\hbar}\right]\psi(\rvec)
- \half{\cal D}(\rvec)\phi(\rvec)\phi(\rvec)\,.\nonumber\\
\eea
The transformation to the classical field theory is effected by positing
that the fields in the equations of motion are no longer boson
fields, but commuting $c$-number fields. The interpretation is that
$\phi(\rvec)$ and $\psi(\rvec)$ are the macroscopic wave functions for atomic and
molecular condensates.

 From this point onward we again take the driving field to be a simple plane
wave with the positive frequency part $\half\E e^{i\q\cdot\rvec}$. Second,
we scale the atomic and molecular fields with the square root of  density,
$\sqrt{N/V}$. Third, we incorporate the spatial variation of the electric
field into the definition of the molecular field. Altogether, we define new
atomic and molecular fields $\Phi$ and $\Psi$ as
\beq
\phi = \sqrt{N\over V}\,\Phi,\qquad \psi =
\sqrt{N\over V}\,e^{i\q\cdot\rvec}\Psi\,.
\eeq
The net results are four. First, the normalization of the fields now reads
\beq
{1\over V}\int d^3r\,(|\Phi|^2 + 2|\Psi|^2) = 1\,.
\label{NOR}
\eeq
Second, the coefficient $\cal D$ gets multiplied by the square root of the
atom density that would prevail if all molecules were to dissociate. Third,
all explicit position dependence disappears from the equations of motion.
The GPE for the rescaled atomic and molecular wave functions are
\bml
\bea
i\dot\Phi(\rvec) &=&\left[-{\hbar\nabla^2\over 2m}-\mu\right]\Phi(\rvec)
-e^{-i\varphi}\Omega\,\Phi^*(\rvec)\Psi(\rvec),
\label{AEQ}\\
i\dot\Psi(\rvec) &=&\left[-{\hbar\over 4m}\nabla^2 + i{\hbar\over
2m}\q\cdot\nabla+
\delta-2{\mu\over\hbar}\right]\Psi(\rvec) \nonumber\\&&-
\half e^{i\varphi}\Omega\,[\Phi(\rvec)]^2\,.
\label{MEQ}
\eea
\label{EQNS}
\eml
None other than our characteristic frequency scale for 
photoassociation, $\Omega$ of
Eq.~\eq{OME}, is now explicitly the frequency scale in the field
equations as well. The phase factor $e^{i\varphi}$ accounts for the phases
of the electric field and dipole moment,  and will soon prove
inconsequential. Fourth, a photon recoil term got added to the
kinetic energy of the molecules.

While still using full dimensional units, we pause to discuss the
mathematical symmetries of the GPE. First, a trivial phase change of one
of the fields, e.g., $\Phi\rightarrow e^{-i\varphi/2}\Phi$, converts
Eqs.~(\ref{EQNS}) to the same equations, except that the phase factors
vanish from atom-molecule interaction terms. Therefore, we drop the phase
in the interaction term. Second, the GPE have a global gauge invariance.
If the fields
$\Phi$,
$\Psi$ are a solution, then so are the fields
$\Phi e^{i\varphi}$, $\Psi e^{2i\varphi}$ for
an arbitrary fixed phase $\varphi$. In particular, putting $\varphi= 2\pi$,
one may see that the equations are invariant under the change of the sign
of the field $\Phi$. Third, as a time dependent generalization of the gauge
invariance, the GPE is invariant under the replacement
\bml
\bea
\Phi&\rightarrow&\Phi e^{i\Delta\mu t},\\
\Psi&\rightarrow&\Psi e^{2i\Delta\mu t},\\
\mu&\rightarrow&\mu+\Delta\mu\,.
\eea
\label{MTR}
\eml
  Fourth, the GPE is Galilei invariant, in
that the replacements
\bml
\bea
\Phi(\rvec,t)&\rightarrow&
e^{i\kvec\cdot\rvec}\Phi\left(\rvec-{\hbar\kvec\over m}t,t\right),\\
\Psi(\rvec,t)&\rightarrow&
e^{2i\kvec\cdot\rvec}\Psi\left(\rvec-{\hbar\kvec\over m}t,t\right),\\
\mu&\rightarrow&\mu + {\hbar^2\kvec^2\over 2m}\,,\\
\delta&\rightarrow&\delta - {\displaystyle\hbar\kvec\over\displaystyle
m}\cdot \q\label{DOPSHIFT}
\eea
\label{GAL}
\eml
convert a solution into another solution that corresponds to an added
momentum $\hbar\kvec$ per atom. The nontrivial part of the
transformation, Eq.~\eq{DOPSHIFT} is that,  due to the Doppler shift, the
effective detuning changes depending on the overall motion of the
atom-molecule system.

Finally, we express the GPE~\eq{EQNS} in a specific system of units,
\beq
t_0 = {1\over\Omega},\quad r_0 = \sqrt{\hbar\over m\Omega},\quad m_0=m\,,
\eeq
for time, length and mass respectively. Technically, we should rename the
scaled variables and parameter; say, $t=\bar{t}\,t_0$,
$\delta=\bar\delta/t_0$. However, we eschew
such a heavy notation, and continue to use, e.g., the symbol $\delta$ for
what now stands for a dimensionless number and should be properly denoted
by $\bar\delta = \delta/\Omega$. The result is
\bml
\bea
i\dot\Phi &=&\left[-\half\nabla^2-\mu\right]\Phi
-\Phi^*\Psi,
\label{DAEQ}\\
i\dot\Psi &=&\left[-\quarter\nabla^2 + \half i\q\cdot\nabla+
\delta-2\mu\right]\Psi -
\half \Phi^2\,.
\label{DMEQ}
\eea
\label{DLGPE}
\eml

The GPE equations are as stated in Ref.~\cite{JAV99a}. The plan now is to
analyze them. Of course, equations of the type~\eq{DLGPE} are standard fare
in studies of second-harmonic generation. The literature once more is
extensive, and we will only give a few specific pointers as we go along.

\subsection{Steady state of GPE}

Atomic and molecular fields $\Phi$ and $\Psi$ represent a stationary
state, one in which the physics does not change with time, if and only if
their time evolution is solely in global (position independent) phase
factors, possibly different ones for $\Phi$ and $\Psi$. Now consider
fields of the form
$\Phi(\rvec,t)=\Phi(\rvec)e^{-i\omega t}$ and $\Psi(\rvec,t)=\Psi(\rvec)e^{-2i\omega t}$
for any real $\omega$. One sees right away that for this type of
evolution, at least the exponential time dependence properly matches on both
sides of Eqs.~\eq{DLGPE}. Although we have not been able to prove it
mathematically, we conjecture that, assuming time  independent parameters
in Eqs.~\eq{DLGPE}, this type of time dependence is also the only possible
case in which the physics is independent of time.

But then, by virtue of the transformation~\eq{MTR}, by readjusting the
chemical potential one can always remove the time dependence of the fields
entirely in any stationary state.  The chemical potential started its
life as an arbitrary parameter. From now on we make use of the
arbitrariness and choose the value in such a way that in steady state the
atomic and molecular fields are literally constants in time.

We shall see shortly that, for any time independent detuning
$\delta$, one may pick a suitable value for the chemical
potential $\mu$ so that the GPE~\eq{DLGPE} have solutions $\Phi_0$, $\Psi_0$
that are constants in both space and time. But by virtue of the Galilean
transformation~\eq{GAL}, we may then construct from $\Phi_0$, $\Psi_0$ a
stationary solution corresponding to an arbitrary overall flow of atoms and
molecules, solutions of the form $\Phi\propto e^{i\kvec\cdot\rvec}$ and
$\Psi\propto e^{2i\kvec\cdot\rvec}$. Conversely, we believe that all
spatially homogeneous stationary solutions are such Galilean boosts of the
once-and-for-all constant solutions $\Phi_0$, $\Psi_0$. As far as it
comes to spatially homogeneous steady states, we therefore may, and will,
without any restriction on generality consider fields that are constants in
space and time.

In the context of second-harmonic generation it is well known that the GPE
may have spatially inhomogeneous solitary-wave type stationary
solutions~\cite{HE96}. By applying the Galilean transformation, one may
find traveling solitary waves as well. Here we will make no effort to
find, let alone classify, solitary-wave solutions to our GPE, but proceed
as if the homogeneous steady states were all there is to it.

Within the scope of the present paper, the stationary solutions thus satisfy
\bml
\bea
0 &=&-\mu\Phi_0
-\Phi^*_0\Psi_0,
\label{E1}\\
0 &=& (\delta-2\mu)\Psi_0 - \half \Phi_0^2\\
1 &=& |\Phi_0|^2 + 2|\Psi_0|^2\,,\label{FLDNOR}
\eea
\label{SSEQ}
\eml
where the final equation originates from the normalization~\eq{NOR}.
Moreover, by virtue of the global gauge invariance, one may choose $\Phi_0$
real. Next, because of the invariance of the GPE with respect to the sign
change of $\Phi$, we may always choose $\Phi_0$ to be nonnegative.
Furthermore, if $\Phi_0>0$, then~\eq{E1} shows that $\Psi$ (like $\mu$) must
be real. On the other hand, if $\Phi_0=0$, $\Psi_0$ comes with an arbitrary
phase, and may be chosen real. All told, we only need to find the
real solutions $\Phi_0$, $\Psi_0$, $\mu$ to Eqs.~\eq{SSEQ}, and besides
only solutions with
$\Phi_0\ge0$ need be retained.

There are three distinct solutions. The trivial one reads
\beq
\Phi_0^0=0,\quad\Psi_0^0=\hbox{$1\over\sqrt{2}$},\quad\mu^0 = \half\delta\,;
\label{MU0}
\eeq
for $\delta\le\sqrt{2}$ we have
\bea
\Phi_0^+ &=& {\sqrt{6-\delta^2-\delta\sqrt{6+\delta^2}}\over3},\nonumber\\
\Psi_0^+ &=& {-\delta-\sqrt{6+\delta^2}\over6},\nonumber\\
\mu^+ &=& -\Psi_0^+\,;
\label{MUP}
\eea
and for $\delta\ge-\sqrt{2}$ we find
\bea
\Phi_0^- &=& {\sqrt{6-\delta^2+\delta\sqrt{6+\delta^2}}\over3},\nonumber\\
\Psi_0^- &=& {-\delta+\sqrt{6+\delta^2}\over6},\nonumber\\
\mu^- &=& -\Psi_0^-\,.
\label{PSIM}
\eea

At least two steady states are found for each $\delta$, and three in the
interval $-\sqrt{2}<\delta<\sqrt{2}$. The question is, which one represents
the desired physical steady state. Here we attempt to mimic the ground
state of the quantum-mechanical two-mode model~\eq{2MH}, and choose the
stationary solution so that~\cite{JAV99a}
\bea
\delta<-\sqrt{2}&:& \Phi_0=\Phi_0^0, \Psi_0=\psi_0^0,\mu=\mu^0,\nonumber\\
\delta\ge-\sqrt{2}&:& \Phi_0=\Phi_0^-, \Psi_0=\psi_0^-,\mu=\mu^-\,.
\label{APPSOL}
\eea
The success is evident in Fig.~\ref{MOLFRA}, where we plot side by side the
fraction of atoms converted to molecules from the quantum-mechanical
two-mode model and the corresponding GPE approximation $f=2\Psi_0^2$ (solid
line) as a function of detuning. By comparing with the $N=10$ and $N=100$
quantum results, it may be seen that the agreement between the GPE and the
quantum approach gets better as the number of atoms is increased.
Even for an atom number as small as 100 and around the
nonanalytic point $\delta=-\sqrt{2}$ of the GPE
approximation~(\ref{APPSOL}), the difference is only on the order of one
per cent.

We conclude by noting that Eqs.~\eq{MU0} and~\eq{MUP} together give a
stationary solution that is the exact mirror image of our
choice~\eq{APPSOL} with the substitution $\delta\rightarrow-\delta$. This
corresponds to the state with maximum energy for the two-mode quantum
system. It is a stationary state just as is the minimum, and may be used
for atom-molecule conversion. The difference is that,
in the case of the maximum, the detuning would be swept in the opposite
direction, from below to above the dissociation threshold. Otherwise
rapid adiabatic passage should work essentially as before.

\subsection{Small excitations of the system}

So far, while analyzing the GPE, we have achieved nothing more than in our
studies of the two-mode model; rather less, because in our classical field
theory we lose quantum fluctuations. Nevertheless, we now have the tools to
analyze the stability of the steady state, and see how spatial
inhomogeneities evolve in time.

As in Ref.~\cite{JAV99a}, we linearize the GPE around a stationary
solution, and then attempt to find eigenmodes for small deviations from the
stationary case. Both of these steps are achieved at once by inserting
the Ansatz
\beq
\Phi(\rvec,t) = \Phi_0 + u_\Phi\,e^{i(\p\cdot\rvec-\omega t)} + v^*_\Phi\,
e^{-i(\p\cdot\rvec-\omega^* t)}
\eeq
into the GPE, and only retaining the first-order terms in the ``small''
coefficients $u_\Phi$, $v_\Phi$.  Of course, the field $\Psi$ is treated
similarly. We need to mix plane waves $\p$ and $-\p$ because the GPE mixes
fields and their complex conjugates. With our Ansatz we also retain the
possibility that the evolution frequency of an eigenmode could be complex.

The Ansatz succeeds if the as of yet unknown evolution frequency
$\omega$ satisfies the eigenvalue equations
\bml
\bea
{\left[
-\half{\p}^2+\mu+\omega
\right]{u}_\Phi +
\psi_0{v}_\Phi + \Phi_0{u}_\Psi=0,\quad}&&\\
{\left[
-\half{\p}^2+\mu-\omega
\right]{v}_\Phi +
\psi_0{u}_\Phi + \Phi_0{v}_\Psi =0,\quad}&&\\
{\left[
-\quarter(\p^2+2\p\cdot\q) + 2\mu-{\bar\delta}+\omega
\right]{u}_\Psi + \Phi_0{u}_\Phi =0,\quad}&&\\
{\left[
-\quarter(\p^2-2\p\cdot\q) + 2\mu-{\bar\delta}-\omega
\right]{v}_\Psi + \Phi_0{v}_\Phi =0.\quad}&&
\eea
\label{EVP}
\eml
The characteristic equation is fourth order in $\omega$, so in principle the
solutions can always be written down analytically. However, here we use
{\it Mathematica\/}~\cite{MATHEMATICA} to produce results directly
numerically, and occasionally to extract analytical forms for special and
limiting cases.

The dispersion relations of the excitation modes, $\omega = \omega(\p)$,
depend on the relative propagation directions of the excitation and of
light, and also on the size of the photon recoil kick. We encompass these
dependences into a dimensionless parameter, which in terms of the original
{\em dimensional\/} parameters reads
\beq
\eta = \sqrt{\hbar(\q\cdot\p)^2\over m\Omega\,\p\cdot\p}\,.
\label{XIDEF}
\eeq
Given the parameter $\eta$ and the absolute value $p$ of the
(dimensionless) propagation vector $\p$, the terms that depend on photon
recoil in Eqs.~\eq{EVP} are replaced as follows,
\beq
\quarter(\p^2\pm2\p\cdot\q)\rightarrow\quarter(p^2\pm2\eta p)\,.
\eeq

For fixed values of the parameters $\delta$, $\eta$ and $p$, there are four
excitation modes with four (in general) different evolution frequencies
$\omega$. If a positive imaginary part is encountered in any one of the four
frequencies, the corresponding mode grows exponentially and the steady
state is unstable.

Let us take the wave vector characterizing the
small-excitation mode to be perpendicular to
the propagation direction of light,
$\eta=0$. We begin with
$p=0$ and assume
$|\Psi_0|^2<\half$, a nontrivial mix of atoms and molecules. We find
the solutions to the characteristic equation for the eigenvalue
problem~\eq{EVP}
\bml
\bea
\omega_1^2(0) &=&0.\\
\omega_2^2(0) &=&\left(3 + {1\over2|\Psi_0|^2} \right)
\left({1\over2}-|\Psi_0|^2\right)\,.
\eea
\eml
When the wave number of the excitation moves away from $p=0$, of the four
frequencies the two evolving continuously from $\pm\omega_2(0)$ remain real.
Their dispersion relations for small $p$ are of the type
$\omega_2(p)\simeq\omega_2(0) + {1\over 2m^*}p^2$, with $m^*\ge 0$, so
that these excitations are akin to optical phonons. On the other hand, the
remaining two excitation frequencies are either purely real or purely
imaginary, and for small enough $p$ they are always imaginary. Such modes
do not propagate at all, but grow or shrink in place exponentially.

\begin{figure}
\includegraphics[scale=.50]{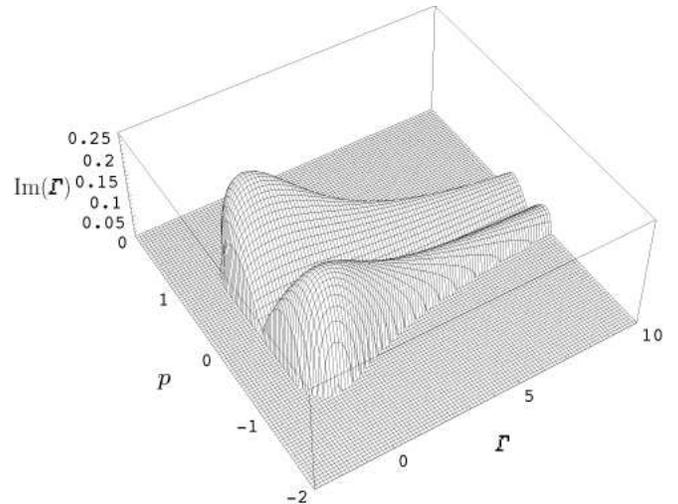}
\caption{Largest imaginary part among the four evolution frequencies
of small perturbations of the joint atom-molecule condensate is plotted as
a function of the laser detuning $\delta$ and the wave number of the
plane-wave like perturbation $p$. In this figure the excitation mode
propagates perpendicularly to the propagation direction of light, so that
$\xi=0$.}
\label{IMP}
\end{figure}

In fact, in Fig.~\ref{IMP}, which is transcribed from Ref.~\cite{JAV99a}, we
plot the largest imaginary part among the four frequencies
$\omega$, $|\Im[\omega_1]|$, as a
function of the detuning
$\delta$ and the wave number
$p$ of the excitation mode. As above, we set $\eta=0$. It
may be seen that, for any detuning
$\delta>-\sqrt{2}$, an unstable excitation mode is always found. The
largest imaginary part of an evolution frequency, i.e, the largest growth
rate of an instability, is encountered at
$\delta=-0.154496$ and $p=\pm0.771324$, and equals $\Im(\omega) = 0.24256$.

The analogous instability is naturally known in second-harmonic generation,
and goes under the rubric ``modulational instability''~\cite{TRI95,HE96a}.

\begin{figure}
\includegraphics[scale=.635]{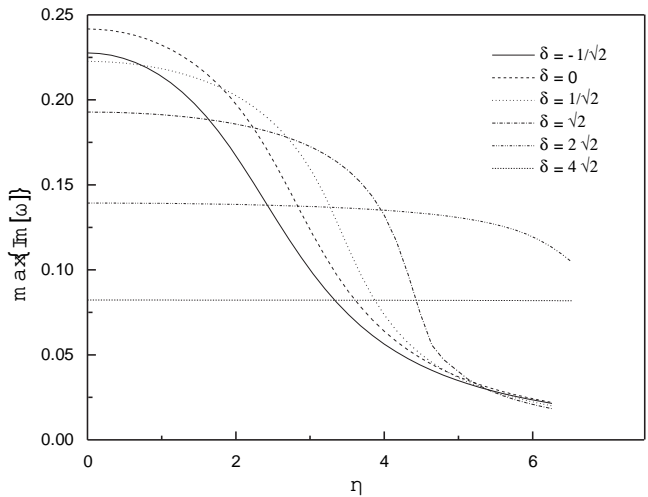}
\caption{Largest imaginary part in a small-excitation evolution frequency
encountered for any value of the wave number of the excitation $p$ is
plotted against the parameter $\xi$ that characterizes the component of
the photon recoil momentum in the direction of the propagation vector of the
small excitation $\p$. The explicit definition of $\xi$ is given in
Eq.~(\protect\ref{XIDEF}). Different curves are for different fixed
detunings $\delta$, as identified in the legend.}
\label{IMPXI}
\end{figure}

We next come to the effect of the direction of the wave vector of the mode
on instability. In Fig.~\ref{IMPXI} we plot the largest positive imaginary
part of a mode frequency found for any $p$ as a function of the parameter
$\eta$. The curves are labeled with their corresponding fixed detunings
$\delta$. In our studies of this kind, the largest growth rate of the
instability was always found for $\eta=0$. For a given detuning, the
small excitations whose momentum direction is perpendicular to light
propagation always present the most unstable scenario.

Finally consider the largest growth rate of the instability (among all $\p$)
as a function of the detuning. As may be inferred from Fig.~\ref{IMP},
with increasing detuning $\delta\gsimeq1$ it decreases and occurs at
smaller momenta, i.e., at larger size scales. In fact, for
$\delta\gg1$, the maximum growth rate of an  instability
$R(\delta)=\max\{\Im[\omega(p,\delta)]\}_p$ behaves as
\beq
R(\delta) \sim \left[{1\over2\,\delta} - {5\over4\,\delta^3} + {\cal
O}({1\over\delta^5})\right],
\label{RMAX}
\eeq
and the corresponding position of the maximum $p_M(\delta)$, the value of
the momentum such that $\Im[\omega(p,\delta)]\le \Im[\omega(p_M,\delta)]$
for all $p$, goes like
\beq
p_M(\delta) \sim
\sqrt{1\over\delta}\,\left[1-{1\over\delta^2}+{47\over16\,\delta^4}
+{\cal O}({1\over\delta^6})
\right]\,.
\label{PMAX}
\eeq
Of course, as the largest growth rate of the instability seems to occur
for $\eta=0$, this was our choice in Eqs.~\eq{RMAX} and~\eq{PMAX}.

In sum, we have found that for $\delta\le-\sqrt{2}$ the steady state of
the atom-molecule system is stable, and for any $\delta>-\sqrt{2}$ it is
unstable. Although we have reported only on a specific stationary
solution~\eq{APPSOL}, possibly one out of three, the modes we have not
considered explicitly are all unstable.  But $\delta=-\sqrt{2}$ is also the
watershed, in that below $\delta=-\sqrt{2}$ the steady state is all
molecules ($\Phi_0=0$), and above $\delta=-\sqrt{2}$ some atoms are
involved.  All told, the steady state with everything in molecules is stable
for $\delta\le-\sqrt{2}$, but any steady state involving atoms is
always unstable. The most unstable situation occurs around
$\delta\simeq0$, with about an equal mixture of atoms and molecules. For
$\delta\gg1$ atoms take over, and the time scale for the instability grows
longer.

\subsection{Fate of unstable system}

Linearized stability analysis has shown that a joint atom-molecule
condensate is unstable in the presence of photoassociating light, i.e.,
there are small-excitation modes that grow exponentially. But when a small
deviation from steady state grows exponentially, eventually it is not a
small deviation anymore and the linearized analysis breaks down. To
investigate the fate of the system once the instability has set in, we
integrate the full GPE numerically as in Ref.~\cite{JAV99a}. Our aim is a
qualitative demonstration, so we proceed in $1+1$ dimensions, one spatial
coordinate
$x$ and time
$t$. However, there is nothing in our method that would not immediately
work in higher spatial dimensions. The restriction is merely a matter of
computer time.

Specifically, in our algorithm we discretize a stretch $L$ of the line into
equidistant points $x_i$, and seek to represent the fields at these
discrete points only. For convenience, we use periodic boundary conditions,
so that the value of all functions repeats over $L$. The GPE is integrated
over a time step $h$ in two moves. First we ignore the position derivatives
in the GPE altogether. This implies that the stripped-down version of the
GPE is local; for each $x_i$, $\Phi(x_i)$ and $\Psi(x_i)$ only couple to
$\Phi(x_i)$ and $\Psi(x_i)$. We integrate the local version of the GPE over
the time $h$ separately for each $x_i$ as two coupled differential equations
using a second-order Runge-Kutta step. To prevent a numerical drift of the
norm, we complete the initial step of the algorithm by normalizing the
fields analogously to Eq.~\eq{NOR}. Second, we ignore anything but the
position derivatives in the GPE. The resulting partial version of the GPE is
nonlocal, but in exchange it is linear and does not mix the fields $\Phi$
and $\Psi$. To integrate over the same (sic!) time step as in the initial
part of the algorithm, we first take the Fourier transform of the result
from the first part using the Fast Fourier Transformation (FFT). In Fourier
space position derivatives become local, so, to propagate the fields over
the step $h$, we simply multiply their Fourier transforms by what is now
the local, exact (within FFT), linear time evolution
operator. Finally, in preparation for the next step, we transform back to
real space.

This split-step algorithm is an obvious variation of the time honored
split-operator methods for linear partial differential
equations~\cite{SPLITSTEP}, and has been described before at least by the
group of Firth~\cite{SKR98}. Nonetheless, it comes with a fair dose of
heuristics. It is therefore gratifying that we have been able to verify a
good rate of convergence by using successively smaller time steps to
integrate over a fixed interval of time.

\begin{figure}
\includegraphics[scale=.55]{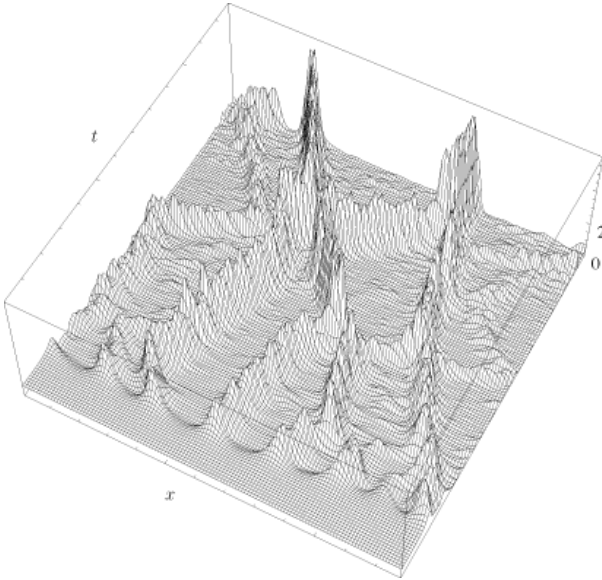}
\caption{An example to demonstrate the modulational instability of the
joint atom-molecule condensate. Atom density $|\Phi|^2$ is plotted as
a function of the spatial coordinate $x$ and time $t$, given that the
system starts out at time $t=0$ in the steady state plus a small amount of
random noise.}
\label{INST}
\end{figure}

We present an example of our simulations in Fig.~\ref{INST}, showing the
absolute square of the atomic field
$|\Phi|^2$ as a function of position $x$ and time $t$. We use 128
points $x_i$. Because of the periodic boundary conditions, the left and
right edges of
$x$ wrap around and are actually the same. The plot is for the parameters
$\delta=0$,
$\eta=0$, the range of position $x$ is 24.6282 (in units of $r_0$)
corresponding to three wavelengths of the most unstable excitation mode for
these parameters, and time $t$ runs from 0 to 127 (in units of $t_0$). The
unit of $|\Phi|^2$, atom density, is such that for a homogeneous gas with
everything in atoms, the density would be $|\Phi|^2\equiv1$. The system
starts at time $t=0$ in the steady state appropriate for $\delta=0$,
$\eta=0$, except that we add a small amount of Gaussian noise to each of the
points specifying the initial state.  Figure~\ref{INST} is otherwise
the same as Fig.~2 in Ref.~\cite{JAV99a}, except that a different seed was
used for the random number generator that added the noise. As befits an
instability, this innocuous change has lead to a totally
different quantitative result at long times.

 From the numerical simulations we see that the nature of the instability is
such that the atoms and the molecules together combine into dense clumps.
These clumps move around, and tend to join when they collide. As far as we
can tell, within the present model only one big, dense clump will remain
in the end.

One might wonder what is the mechanism behind the clumping. We
present a heuristic guess. We are not talking about a thermodynamic system,
so minimization of energy is a dubious principle to begin with; and
besides, we are dealing with quasienergies in a rotating frame, not real
energies. With these warnings out of the way, let us surmise
that the system nevertheless attempts to minimize its energy. The atomic and
molecular condensates make something akin to the two-level system in
quantum optics. When light is added, one gets a dressed two-level system.
The energy of the lower one of the two dressed states decreases with
increasing Rabi frequency, which is the product of the electric field
strength of the laser and the dipole moment. But the analog of Rabi
frequency for the two-level system of atomic and molecular condensates,
$\Omega$, is also proportional to the square root of density, so the present
system may also decrease its energy by increasing the density. Maybe this
is what the instability is about.

\section{Experimental considerations}\label{EXCON}

The models we have discussed until now have been rather rudimentary. We next
take up two types of complications that may come up in experiments. First,
in Sec.~\ref{ROGUE} enters an angle that is in principle included in our
many-body Hamiltonian, although we have not yet considered it expressly:
photodissociation of condensate molecules to states outside of the atomic
condensate. It turns out to limit the rate at which one can achieve
coherent atom-molecule conversion in adiabatic passage. Second, in
Sec.~\ref{MISSING} we discuss a number of aspects that have so
far been missing from our models: spontaneous emission
from the photoassociated state, atom-atom interactions, trapping of atoms
and molecules, and various level shifts. Spontaneous emission can be
ameliorated by resorting to a two-color photoassociation scheme. Provided
that photoassociation is speedy enough, which we believe is technically
possible, the rest of these complications may be minor nuisances rather
than dominant features of an experiment. What it takes to make
photoassociation speedy enough is the subject of Sec.~\ref{EXPAR}, where we
discuss the characteristic Rabi frequency for photoassociation $\Omega$ for
various alkalis. Finally, in Sec.~\ref{HEIZEXP}, we analyze the published
experiment of Ref.~\cite{WYN00} from our viewpoint of coherent
photoassociation.

\subsection{Rogue photodissociation}\label{ROGUE}

As we noted already, in the case of photoassociation of an infinite homogeneous
condensate, momentum conservation uniquely determines the state of the
ensuing molecule. The converse, however, does not hold. When a molecule
photodissociates into two atoms, momentum conservation does not force the
atoms to return to the atomic condensate. Bose enhancement favors
recombination of atoms with the condensate; the characteristic frequency
is $\Omega$ for both photoassociation {\em and\/} photodissociation 
between atomic and molecular
condensates. Nonetheless, atoms winding up elsewhere are lost for coherent
photoassociation. We have coined the term ``rogue photodissociation'' 
for photodissociation processes
that send atoms outside the atomic condensate.

One might think that energy conservation in the cycle of 
photoassociation and photodissociation is the
additional constraint that guarantees that the atoms return to the
condensate. But this need not be a compelling argument. Any time
dependence in the system interferes with energy conservation. For instance,
suppose that photodissociation proceeds to the noncondensate states at the
same rate $\Gamma_0$ (per atom) as it would in the case of a nondegenerate
gas of molecules, so that after a time $\sim\Gamma_0^{-1}$ coherent 
photoassociation
ceases. The photodissociation rate $\Gamma_0$ in itself furnishes a time  
scale such that
energy has to be conserved only to within $\hbar\Gamma_0$. The time
evolution involved in nonlinear Rabi flopping would also interfere with
energy conservation.

We present here a rudimentary model for rogue photodissociation for the
special case when the detuning is swept in order to convert an atomic
condensate to a molecular condensate. The key assumption is that we may
employ the standard Markov approximation in the analysis of
photodissociation. This entails that rogue photodissociation has no
memory, but is characterized at each instant of time by a rate of
exponential decay. Such an assumption seems dubious in particular when
the laser is tuned to the close vicinity of the photodissociation 
threshold~\cite{RZA82}. However, we know of no near-threshold case of
this kind in which the breakdown of the Markov approximation has proven
relevant in an experiment.

Evidently, only the condensate mode, one of very many atomic modes, is
strongly affected by Bose enhancement. We take rogue photodissociation 
to proceed at the rate that would be appropriate for a nondegenerate gas
at the given detuning. Second, we model the dependence of the
photodissociation  rate on detuning using
the Wigner threshold law, so that we write
\beq
\Gamma(\delta) =
\theta(\delta)\,\sqrt{\delta\over\Omega}\,\Gamma_0\,.
\eeq
For convenience we have chosen the photoassociation frequency scale 
$\Omega$ as the
reference detuning for photodissociation rate; $\Gamma_0$ is the 
photodissociation rate
for the detuning $\delta=\Omega$. As the third quantitative
element of the model, we take the probability that a given atom is in the
molecular condensate to be twice the square of the molecular field amplitude
as solved from the classical field theory, and normalized as in~\eq{FLDNOR}.
Explicitly, in dimensional units and for
$\delta\ge0$, this probability is found from Eq.~\eq{PSIM} as
\beq
P_M(\delta) = {1\over 18} \left[-\left({\delta\over\Omega}\right)
+\sqrt{6+\left({\delta\over\Omega}\right)^2} \right]^2
\eeq
Suppose now that the detuning is swept as $\delta=-\xi\Omega^2t$, as
in our rapid adiabatic passage example. Ignoring the depletion of the
condensates due to the very same rogue photodissociation, we find the 
total probability for
rogue photodissociation
\bea
P &=& \int_{\delta(t)\ge0} dt\, P_M[\delta(t)]
{\,\Gamma_0\sqrt{\delta(t)\over\Omega}}\nonumber\\
&=&{\Gamma_0\over\xi\Omega}\left[
{2^{3/4}\over3^{1/4}}\int_0^\infty
d\tau\,\sqrt{\tau}\left(\sqrt{1+\tau^2}-\tau\right)^2
\right]\nonumber\\
&=&\alpha\,{\Gamma_0\over\xi\Omega}\,,
\label{RPFR}
\eea
where the numerical constant has the value $\alpha=4.03197$. Except for the
numerical factor, this is the same expression we already used in a
qualitative estimate in Ref.~\cite{JAV99}.

Since the photodissociation rate grows linearly with light intensity,
$\propto I$, and  the photoassociation characteristic
frequency $\Omega$ is proportional to the field strength of the laser,
$\propto\sqrt{I}$, in the end rogue photodissociation wins out as the
light intensity is increased. Qualitatively, when $P=1$, rogue
photodissociation has overtaken coherent conversion of atoms to
molecules. To study this borderline case, we first set
$P=1$ in Eq.~\eq{RPFR} and solve $\Gamma_0$ as a function of
$\Omega$. We then insert the result into Eq.~\eq{OME}, thus eliminating
$\Gamma_0$. Moreover, the velocity $v_0$ in Eq.~\eq{OME}
is then the  relative velocity of the dissociated atoms corresponding to the
detuning of the laser that gave the photoassociation rate $\Gamma_0$, in
this case
$\delta=\Omega$. Therefore we have $v_0^2/2\mu=\hbar\Omega$, and $v_0$ may
be eliminated as well. We finally solve for the borderline value $\Omega$
as a function of the problem parameters. After simple manipulations the
ensuing characteristic frequency scale for photoassociation may be written
\beq
\hat\Omega =
\left({8\sqrt{2}\pi\xi\over\alpha}\right)^{2/3}\,(\rho\lambdabar^3)^{2/3}
\,\epsilon_R\,.
\label{MAXFRQ}
\eeq
Here we have introduced $\lambdabar$, wavelength of the light
divided by $2\pi$, and the familiar recoil frequency for laser cooling
\beq
\epsilon_R = {\hbar\over 2m\lambdabar^2}\,.
\eeq

The main finding is that rogue photodissociation restricts the light
intensity that one may profitably use for adiabatic atom-molecule
conversion. This means that there is also a maximum usable
photoassociation frequency, or a minimum possible time scale for adiabatic
atom-molecule conversion. The way we have written our estimate~\eq{MAXFRQ},
the frequency scale is provided by the photon recoil frequency, and the
corresponding time scale is in the ballpark of
$\epsilon_R^{-1}$. The density dependence of photoassociation is
encapsulated in the parameter
$\rho\lambdabar^3$, the usual dimensionless parameter that governs the
coupling of light with matter in a dense medium. For present-day
condensates
$\rho\lambdabar^3\sim~1$ is a reasonable rule of thumb. Finally,
we have a numerical constant that depends on the rate of sweeping of the
detuning, but which may also be set equal to one in a rough estimate.
Altogether, when in need of a qualitative number for the 
photoassociation frequency
$\Omega$, we resort to $\Omega\sim\epsilon_R$.

Although our estimate of the minimum time scale for coherent
photoassociation was developed for rapid adiabatic passage, we believe
that (with $\xi\simeq1$) it also applies to Rabi flopping. This is
because the dimensional parameters of the problem are the same in both
cases. In fact, as it comes to dimensional quantities, the minimum time
scale for coherent photoassociation is equivalently written~\cite{FOOT}
\beq
\tau\sim{m\over\hbar\rho^{2/3}}\,.
\label{MINTIME}
\eeq
This is the essentially unique quantity with the dimension of time that
can be put together using the quantities characterizing a homogeneous,
noninteracting, quantum mechanical, zero-temperature BEC; density, atom
mass, and
$\hbar$.

\subsection{Physics missing from model}\label{MISSING}

\subsubsection{Spontaneous emission}\label{CONSGAMMA}

We have discussed a one-color model for photoassociation. The physical
drawback is that,
where there is a strong dipole matrix element for
photoassociation/dissociation driven by external light, there is also a
strong dipole matrix element for spontaneous emission. For instance, if a
photon is absorbed in photoassociation, then there is also a reverse
spontaneous decay of the molecule. The  molecule may end up in bound
vibrational states, either in the same electronic manifold where the atoms
started from, or in some other electronic manifold. Alternatively, the
photoassociated molecule may decay back to a dissociation continuum in a
process known as radiative escape. Either way, usually the probability is
small that the system returns to the same two-atom state in which is
started. After each process of spontaneous emission, two atoms are
typically lost for any profitable use.

The analogous problem of an unstable excited state is standard fare in
quantum optics, and so is the solution: add another laser-driven transition
from the spontaneously decaying state to a stable state. In the same way,
two-color Raman photoassociation, a free-bound transition followed by 
a bound-bound
transition of the molecule, may take place between (nearly) non-decaying
atomic and molecular states~\cite{BOH96,MAC99,BOH99}. Here we study Raman
photoassociation as a means of achieving an effective two-mode scheme;
genuine three-mode phenomena such as STIRAP are discussed
elsewhere~\cite{MAC00}

\begin{figure}
\begin{center}
\includegraphics[scale=.55]{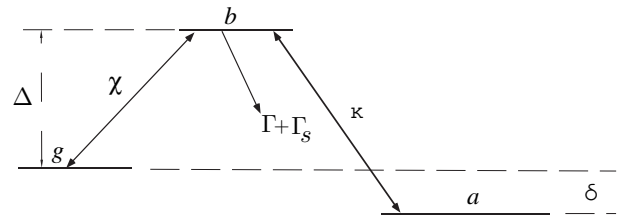}
\end{center}
\caption{Schematic three-level scheme for photoassociation. The levels
$a$, $b$ and $g$ stand for atoms, primary photoassociated
molecules, and final stable molecules, respectively. The figure is drawn
in the rotating frame, so that the Rabi frequencies $\kappa$ and
$\chi$ representing the laser fields are independent of time, and the
intermediate and two-photon detunings $\Delta$ and $\delta$ are drawn as
appropriate level energies.}
\label{3LS}
\end{figure}

Our present notation for this scheme is sketched in Fig.~\ref{3LS}. Suppose
the first step of photoassociation takes place with absorption of a photon,
then one sets up another laser to force, say, induced emission from the
primary photoassociated state to a (more) stable bound molecular state. We
call the Rabi frequency in the second step $\chi$, the detuning of the
laser from resonance in the second transition
$\Delta$, and the spontaneous decay rate of the primary photoassociation
state
$\Gamma_{\rm s}$. In this context $\delta$ stands for the two-photon
detuning, the total energy mismatch for light-induced transition from the
initial atoms to the final stable molecular state, including appropriate
photon recoil corrections.

Let us model our three-level $\Lambda$ scheme using a variation of the
two-mode model~\eq{2MH} as
\beq
{H\over\hbar}=\delta\,g^\dagger g
+(\Delta\!+\!\delta) b^\dagger b -\chi(b^\dagger g\!+\! g^\dagger b)-
\half\kappa (b^\dagger a a \!+\! b a^\dagger a^\dagger)\,,
\label{3MH}
\eeq
where $g$, for ``ground'', is the annihilation operator
for the stable molecular state. Once more, we have added a
multiple of the conserved particle number, this time in such a way that
the atoms have zero energy. The term $\propto\chi$ describes transitions
between the bound molecular states. Spontaneous losses from the
intermediate state are ignored for the time being.

The Heisenberg equations of motion from Hamiltonian~\eq{3MH} are
\bml
\bea
\dot{g} &=& -i\delta g + i\chi b,\label{GDOT}\\
\dot{b} &=& -i(\Delta+\delta) b + i \chi g + \half i\kappa a
a,\label{BDOT}\\
\dot{a} &=& i\kappa a^\dagger  b\label{ADOT}\,.
\eea
\eml
We next eliminate the intermediate state adiabatically with the
assumption that $\Delta$ is the largest evolution frequency in the system.
We thus formally set $\dot{b}=0$, and obtain
\beq
b \simeq{\chi g + \half \kappa a a\over\Delta}\,.
\label{ADE}
\eeq
Inserting this into Eqs.~\eq{GDOT} and~\eq{ADOT}, we have
\bea
\dot{g} &=& -i \left(\delta-{\chi^2\over\Delta}\right)g+\half
i\,{\kappa\chi\over\Delta}\, aa,\label{HEQ1}\\
\dot{a} &=&i\,{\kappa\chi\over\Delta} a^\dagger g + \half i
{\kappa^2\over\Delta} a^\dagger a a\,.\label{HEQ2}
\eea
These are Heisenberg equations of motion for the effective Hamiltonian
\beq
{H_{\rm eff}\over\hbar} = \left(\delta-{\chi^2\over\Delta}\right)g^\dagger g
-\quarter {\kappa^2\over\Delta} a^\dagger a^\dagger a a - \half
{\kappa\chi\over\Delta}(a^\dagger a^\dagger g + g^\dagger a a)\,.
\eeq

The effective Hamiltonian describes a two-level system with the two-photon
detuning $\delta$ and two-photon Rabi frequency $\kappa\chi/\Delta$ in lieu
of the usual detuning and Rabi frequency. There are two additional twists
to the story. First, the two-photon resonance experiences a light shift
$-\chi^2/\Delta$, an old acquaintance from quantum optics. Second, we
have an effective atom-atom interaction proportional to
$\kappa^2/\Delta$. An analogous interaction in a nondegenerate gas
was discussed earlier in Ref.~\cite{SHL96}. Other than these tweaks,
everything we have said about the two-mode model applies as before.

The adiabatic elimination~\eq{ADE} has a dark side hidden by a notational
trick, namely, that it does not preserve boson commutators. Had we written
the right-hand side in Eq.~\eq{ADOT} as $b a^\dagger$ instead of $a^\dagger
b$ and substituted~\eq{ADE}, the atom-atom interaction term in the
effective Hamiltonian would have displayed the operator ordering
$aaa^\dagger a^\dagger$ instead of $a^\dagger a^\dagger a a$. The
difference is immaterial for a large atom number, the limit we are studying
anyway, but a more careful investigation of the adiabatic elimination would
be in order if the atom number were not large.

As excessive care with operator products is not warranted, we
write from the adiabatic assumption the number of atoms in the
intermediate state $b$ qualitatively as
\beq
b^\dagger b \sim \left({\chi\over\Delta}\right)^2 g^\dagger g +
\left({\kappa\over 2\Delta }\right)^2 (a^\dagger a)^2\,.
\label{MIP}
\eeq
Let us scale the operators by $\sqrt{N}$,
i.e., write $b=\sqrt{N}\beta$, and so forth. Then, within a factor of two,
the quantum expectation value $\langle\beta^\dagger\beta\rangle$ is the
probability that an atom is in the intermediate state, and so on.
Equation~\eq{MIP} becomes
\beq
\beta^\dagger\beta \sim \left({\chi\over\Delta}\right)^2
\gamma^\dagger\gamma +
\left({\Omega\over 2\Delta }\right)^2 (\alpha^\dagger\alpha)^2\,.
\label{IMFRAC}
\eeq
The fraction of atoms lost per unit time to spontaneous emission from the
intermediate state equals $2\Gamma_{\rm
s}\langle\beta^\dagger\beta\rangle$.The intermediate detuning $\Delta$
suppresses losses by a factor $\propto1/\Delta^2$, whereas the effective
Rabi frequency scales as $1/\Delta$. In principle, and at the present
level of the physical model, it is possible to get rid of the harmful
spontaneous emission to any desired degree by increasing the intermediate
detuning.

\subsubsection{Interactions between atoms and molecules}

Atoms interact among themselves, molecules interact with molecules, and
atoms even interact with molecules. For a dilute gas, the atom-atom
interaction is often described by the effective two-body potential
\beq
U(\rvec_1,\rvec_2) = {4\pi\hbar^2 a\over m}\,\delta( \rvec_1-\rvec_2)\,,
\eeq
where $a$ is the $s$-wave scattering length for the atoms as before.
One may write analogous models for atom-molecule and molecule-molecule
interactions. If atom-atom interactions were suspected to be a factor, one
could add  to the field theory the usual two-body atom-atom
interactions as
\beq
{\cal H}_{AA} = {2\pi\hbar^2 a\over m}\,
\phi^\dagger(\rvec)\phi^\dagger(\rvec)\phi(\rvec)\phi(\rvec)\,,
\label{AAI}
\eeq
and so on. Inelastic collisions, such as quenching of the molecules by
collisions, may also prove important. At a phenomenological level, they
could be described by using a complex scattering length for atom-molecule
collisions.

Nonetheless, we have considered neither elastic nor inelastic collisions
explicitly in this paper. The motivation is mainly pragmatic.
Photoassociation is the novelty of this work anyway. Second, at this
point in time virtually nothing is known about the scattering lengths for
cases other than atom-atom interactions. Third, as pointed out in
Sec.~\ref{ROGUE}, we anticipate a characteristic frequency scale for
photoassociation, 
$\Omega$, to be of the
order of photon recoil frequency $\epsilon_R$ of laser cooling, say, ten
kilohertz. This is larger than a typical frequency scale associated with
collisions, $4\pi\hbar a \rho/m$, in many of the present alkali
experiments. Photoassociation should dominate the action at least over
short time scales.

More formally, we see from Eqs.~\eq{MAXFRQ} and~\eq{MINTIME} that the
maximum usable photoassociation frequency scales with atom density as
$\rho^{2/3}$, while the rate of binary collisions scales as $\rho$. In
principle and at this level of modeling, it is always possible to make
photoassociation win out by decreasing the density.

Of course, not all of our discussions are for short times only. Notably, in
the case of the instability of a joint atom-molecule condensate, it may
well happen that collisional interactions eventually play a role in
the clumping. We plan to return to collisional effects in a future
publication, inasmuch as something worthwhile emerges from this front.

\subsubsection{Trapping of atoms}

In the current alkali vapor experiments one does not see infinite
homogeneous condensates, but the condensate is ordinarily confined to a
magnetic trap with a (practically  quadratic) potential $V(\rvec)$. This may
be taken into account in the field theory by adding a term in the
Hamiltonian density,
\beq
{\cal H}_{\rm VA}(\rvec) = \phi^\dagger(\rvec)V(\rvec)\phi(\rvec)\,.
\eeq
A similar additional term would describe trapping of molecules.

In fact, the beauty of the field theoretical formulation is that it does not
depend on any given one-particle basis to describe the motion of the atoms
and molecules. The photoassociation term was originally discussed using
plane-wave states, which makes the derivation easy, but at the level of
field theory there is no manifest vestige of plane waves anymore. Even if
we add trapping potentials, there is no need to tamper with the 
photoassociation term.
This should be contrasted with the more delicate  situation that
emerges if one tries to consider photoassociation directly using the 
eigenstates of the
trap, or indeed some states that would take into account both trapping
and atom-atom interactions.

Once more, if the photoassociation frequency scale is of the order 
$\epsilon_R$,
it is still vastly larger than the typical frequency scales associated with
magnetic trapping of atoms, and the corresponding length scale for
photoassociation, $\lambdabar$, is far smaller than the size of a typical
condensate. Over short times the condensates behave locally as if they were
homogeneous. One just applies the theory of an infinite condensates at each
local density, and averages the results over the trap.

On the other hand, there are cases in our formulation where the trapping
will matter. For $\Omega~\sim\epsilon_R$ the modulational
instability may well {\em set in\/} as in a homogeneous condensate, but the
motion of the atoms and molecules due to the trapping forces will certainly
have a long-term effect on the atom-molecule clumps. We do not discuss this
issue here, but plan to return to trapping in a future publication.

\subsubsection{Level shifts}

We have already mentioned a few mechanisms that can change the position of
the photoassociation resonance. Atom-atom and molecule-molecule 
interactions alter the
energy per atom or per molecule, and thereby modify the resonance condition
for photoassociation. Since the atom-molecule ratio conversely 
depends on the detuning,
the makings of bistability and hysteresis are in principle there. We have
also discussed the light shift and the many-body shift in
a two-color, three-mode configuration.

One more shift we have brought up before~\cite{JAV98,MAC99}, but not yet in
this paper, arises because the dissociation continuum is not flat. Given the
initial state, the dipole matrix element (or more precisely, the square of
the dipole matrix element per unit energy) depends on the final continuum
state. The result is that the photodissociation rate picks up an 
imaginary part. That, of
course, amounts to a shift of the photodissociating state with respect to
the continuum. The shift is proportional to light intensity, and in a
qualitative estimate is comparable to the photodissociation rate; see, e.g.,
Refs.~\cite{jav}.

Moreover, there are additional shifts due to the presence of each and
every discrete state dipole-coupled to the initial state. The sum of all
light shifts is actually finite only because the dipole coupling
eventually tends to zero when one goes high enough in the energy of the
coupled states. The implication is that, {\it a priori}, all dipole
coupled states, even those off resonance by perhaps several photon
energies, have to be considered explicitly. If one finds a significant
contribution to the light shift from {\em one\/} far-off resonance state,
chances are that one has to consider {\em all\/} of them.

We will not attempt to address continuum and non-resonant light shifts
explicitly. Nonetheless, on the basis of the atomic
case discussed in Refs.~\cite{jav}, we believe that, at least above the
photodissociation threshold, the light shift  should be reasonably
independent of where exactly the laser is tuned. This would mean that, in
the rapid adiabatic passage type atom-molecule conversion, the light shift
merely gives a constant bias to the detuning, and is virtually
inconsequential.

\subsection{Numerical examples}\label{EXPAR}

\subsubsection{From rate to Rabi frequency}

One-color photoassociation has been analyzed by various groups
\cite{PA-groups}, in particular for alkalis at finite temperature. A
typical outcome is the photoassociation rate ${\cal R}_{v'}$ (in
s${}^{-1}$) or the photoabsorption rate coefficient $\alpha^{PA}_{v'}$
(in cm$^{5}$) for a given bound level $v'$ of the excited electronic 
state. The latter quantity is independent of two experimental
parameters,  namely atom density $\rho$ and laser intensity $I$. The rate
of photoassociation
${\cal R}_{v'}$ is obtained via
\begin{equation}
   {\cal R}_{v'} = \rho\varphi\alpha^{PA}_{v'} \label{eq:rate} \; ,
\end{equation}
where the photon flux (photons/s\,cm$^{2}$) is given by
$\varphi = I/\hbar\omega_L$, $\omega_L$ being the photon frequency.

But we also know from our earlier work~\cite{JAV98,MAC99} that the
photoassociation rate in a thermal sample is given in terms of our detuning
$\delta$, temperature $T$, and photoassociation rate $\Gamma$ as
\begin{equation}
{\cal R}_{v'} = e^{-{\hbar\delta\over k_B T}}\, \rho
\lambda_D^3\,\Gamma\,,
\label{eq:oldrate}
\end{equation}
where
\beq
\lambda_D = \left({2\pi \hbar^2\over\mu k_B T}\right)^{1/2}
\eeq
  is the usual thermal
de Broglie wavelength, albeit calculated using the reduced mass of the
colliding atoms. In what follows, we write
the detuning parameter for photoassociation as $\delta=
\omega_{\infty}-\omega_{L} -\Delta_{v'}$, where $\hbar\omega_{\infty}$ 
is the asymptotic
energy difference between the electronic curves and $\Delta_{v'}$ is the
red-detuning of level $v'$ from its asymptote.  The binding energy
of the molecular state is thus equal to $\hbar\Delta_{v'}$. Combining
Eqs.~(\ref{eq:rate}) and~(\ref{eq:oldrate}) with Eq.~(\ref{OME}), we have
an expression for the characteristic frequency of coherent photoassociation
to the level $v'$,
\beq
\Omega_{v'} = e^{{\hbar\delta\over 2 k_BT}}\sqrt{2\pi\hbar^2\alpha^{PA}_{v'}
\varphi\rho\over v \mu^2 \lambda_D^3}\,.
\label{eq:OmNum}
\eeq

In all of our discussion below we choose the detuning in such a way that
$\hbar\delta = \half k_B T$, which gives the corresponding resonance
velocity $v = \sqrt{k_B T/\mu}$. Strictly speaking, Eq.~(\ref{OME}) requires
the limit $v\rightarrow0$, or equivalently, $\delta\rightarrow0$. However,
we always assume, without explicitly checking this assumption, that the
temperature is already low enough to bring the system into the region of
validity of the Wigner threshold law. The quotient $\Gamma/v$ in
Eq.~(\ref{OME}) has then supposedly reached the $v\rightarrow0$ limit.

We have already introduced the usual density parameter for light-matter
coupling $\lambdabar^{3}\rho$ to characterize atom density, and the
recoil frequency
$\epsilon_R$ as the frequency scale. In the same vein, we write
the Rabi frequency for photoassociation as
\beq
{\Omega_{v'}\over\epsilon_R} =\sqrt{\left({I\over I_{v'}}\right)}\,
\sqrt{(\lambdabar^3\rho)}\,,
\eeq
where $I_{v'}$ is the characteristic light intensity for coherent
photoassociation. In explicit numbers, we find from Eq.~(\ref{eq:OmNum})
\beq
I_{v'} = {1.46245\times 10^{-26}\over
\left[{\displaystyle m\over\displaystyle{\rm u}}\right]^2
\left[{\displaystyle T\over\displaystyle{\rm K}}\right]
\left[{\displaystyle \lambda\over\displaystyle{\rm nm}}\right]^2
\left[{\displaystyle \alpha^{PA}_{v'}\over\displaystyle{\rm cm^5}}\right]
}\,
{\rm W\over\rm cm^2}\,,
\eeq
which also displays the units used to express atomic mass $m$, temperature
$T$, wavelength of light $\lambda$, and photoabsorption coefficient
$\alpha^{PA}_{v'}$.

\subsubsection{Calculated photoassociation rates}

One can estimate the photoabsorption rate coefficient for a pair of atoms.
At low temperatures, only $s$-wave scattering contributes to the process,
and one finds \cite{COT95,cote-pra}
\begin{equation}
   \alpha^{PA}_{v'}(\omega ,T) \simeq \frac{4\pi^{2}\omega}{3c} \lambda_{D}^{3}
   e^{-{\hbar\delta\over k_{B}T}} |D_{v'}(\hbar\delta)|^{2} \;
,\label{eq:kappa}
\end{equation}
In the low-temperature limit, the dipole matrix element
$|D_{v'}(\hbar\delta)|^{2}$ can be approximated by
\begin{eqnarray}
   |D_{v'}(E)|^{2} & \equiv & |\langle v', J=1|D|\hbar\delta,J=0\rangle \; ,
\nonumber\\
   & \simeq & \left(\frac{2\mu k}{\pi\hbar^{2}}\right) |D_{0}|^{2}
     (a-R_{v'})^{2} S_{v'}^{2}
   \label{eq:dipole}
\end{eqnarray}
with $E=\hbar^{2}k^{2}/2\mu = \hbar\delta$,
where $D_{0}=|\db|$ is the asymptotic dipole moment, $a$ is the scattering
length,
$R_{v'}$ is the classical outer turning point of the excited level $v'$,
and $S_{v'}$
is a dimensionless parameter representing the fraction of the
bound wave function contained in the last node
\cite{COT95,cote-pra,cote-jms}.

For example, given $^{7}$Li atoms in a triplet state at
1 mK, a detailed
calculation \cite{cote-jms} showed that the excited level $v'=58$
has the best Franck-Condon factor with the continuum and the highest-lying
bound level $v''=10$ of the lower triplet electronic state, with
$\alpha^{PA}_{v'=58}=2.0\times 10^{-32}$ cm$^{5}$~\cite{note-rate}.
The corresponding characteristic intensity is $I_{v'=58} = 33\,{\rm
W\,cm^{-2}}$.

\begin{figure}[t]
\includegraphics[scale=.49]{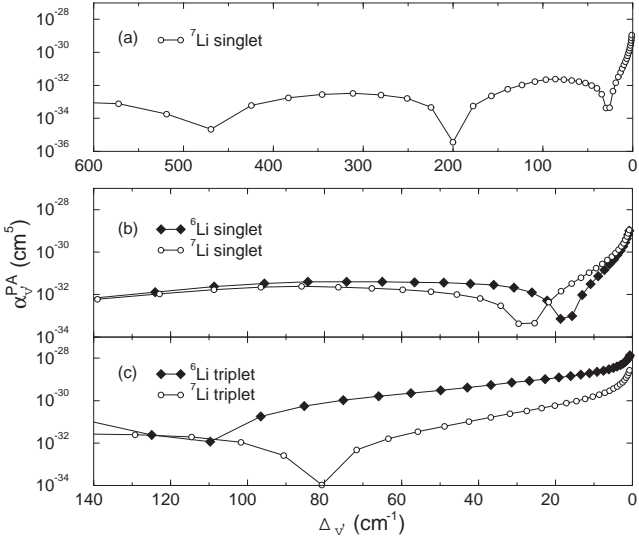}
\caption{Variation of photoassociation rate coefficient 
$\alpha^{PA}_{v'}$ for high-lying vibrational states. In~(a),
the $^{7
}$Li singlet transitions over a range of 600 cm$^{-1}$
exhibit oscillations mapping the nodal structure of the
continuum wave function, with values varying over five orders
of magnitude. Plots (b) and (c) show a shorter range of detunings
for singlet and triplet transitions, respectively. The singlet
scattering lengths of both isotopes are positive and $\alpha^{PA}_{v'}$
decays rapidly from its large values near $\Delta\sim 0$ cm$^{-1}$, 
by nearly
three orders of magnitude. The triplet scattering lengths are both
negative, and $\alpha^{PA}_{v'}$ decreases slowly for small detunings,
until the first nodes of the continuum wave functions are reached.
}
\label{LIRATE}
\end{figure}

Beyond this, in Fig.~\ref{LIRATE} we give the calculated rate coefficients
for high-lying vibrational states for both stable isotopes of Li, for both
the singlet and the triplet excited states. Correspondingly,
Table~\ref{tab1} presents a few numerical examples of the rate
coefficients and characteristic intensities.

  \begin{table}[t]
  \begin{center}
  \caption{
  Photoabsorption rate coefficients $\alpha^{PA}_{v'}$ (in cm$^{5}$)
	and characteristic intensities $I_{v'}$ (in W~cm$^{-2}$)
  for levels $v'$ with the best simultaneous Franck-Condon factors
  with the continuum and the highest-lying level of the lower
  electronic state, and for levels corresponding to detunings near
  1 cm$^{-1}$. The vibrational number $v'$ as well as the
  corresponding binding energy $\Delta_{v'}$ (in cm$^{-1}$) are
  also given. Calculations were performed for T=1 mK.}
  \begin{footnotesize}
  \begin{tabular}{ccccc} \\
    & \multicolumn{2}{c} {\underline{singlet}} &
      \multicolumn{2}{c} {\underline{triplet}} \\
    &  $^{6}$Li & $^{7}$Li &
          $^{6}$Li & $^{7}$Li \\
        &   &   &   & \\ \hline
        &   &   &   & \\
  $v'$ & 70 & 69 & 51 & 58 \\
  $\Delta_{v'}$ & 36.4 & 86.7 & 164 & 122 \\
  $\alpha^{PA}_{v'}$ & $2.8\times 10^{-32}$ & $2.4\times 10^{-32}$ &
                  $2.3\times 10^{-31}$ & $2.0\times 10^{-32}$ \\
  $I_{v'}$ & $32$ & $27$ &
                  $3.9$ & $33$ \\
        &   &   &   & \\
  $v'$ & 87 & 94 & 79 & 86 \\
  $\Delta_{v'}$ & 0.99 & 0.99 & 1.05 & 0.87 \\
  $\alpha^{PA}_{v'}$ & $6.7\times 10^{-30}$ & $7.8\times 10^{-30}$ &
                  $9.2\times 10^{-29}$ & $2.1\times 10^{-29}$ \\
  $I_{v'}$ & $0.13$ & $0.085$ &
                  $0.0098$ & $0.032$ \\
        &   &   &   & \\
  \end{tabular}
  \end{footnotesize}
  \label{tab1}
  \end{center}
  \end{table}

Much larger photoassociation rates and correspondingly smaller characteristic
intensities can be obtained if the excited levels $v'$ are closer to the
dissociation limit (i.e., have smaller binding energies). In fact, the
$v'=58$ level is deeply bound,  with a binding energy of
$\hbar\Delta_{v'=58} = 5.5\times 10^{-4}$ hartree (or 122.22
cm$^{-1}$~\cite{note-hulet}). For example,
$v'=86$ with a binding energy of $0.87\,{\rm cm}^{-1}$ has a rate coefficient
1050 times larger, namely
$\alpha^{PA}_{v'=86}=2.1\times 10^{-29}$ cm$^{5}$, which gives the characteristic
intensity $I_{v'=86} = 0.032\,{\rm W\,cm^{-2}}$. The high-lying excited
levels
$v'$ have larger rate coefficients because the overlap of the excited bound
wave function and the continuum ground wave function is larger. For high
$v'$, the overlap scales like
$\Delta_{v'}^{-1/2}$.

Similar results for sodium and cesium are available.
For small binding energies, the rate coefficient $\alpha^{PA}_{v'}$ at 1 mK
is of the order of $10^{-28}$ cm$^{5}$ for Na, and $10^{-27}$ cm$^{5}$ for
Cs \cite{more-rates}. These translate to characteristic intensities as
small as
$\rm \mu W\,cm^{-2}$. Notice that for lithium,  the rate coefficients for
singlet transitions are smaller than for triplet transitions, reflecting
the sign of the scattering length: this is especially significant for
$^{6}$Li, where the triplet scattering length is negative and enormous
\cite{COT95a}.

Other expressions based on  semi-classical treatment have been developed
\cite{pillet,more-rates}.  For example, in \cite{pillet}, the rate for
$s$-wave contribution is
\begin{equation}
   {\cal R}_{v'} \simeq \rho \lambda_{D}^{3}
   e^{-{\hbar\delta\over k_{B}T}}
\sqrt\frac{\pi^2}{{3\delta\Delta_{v'}}}
\sin^{2}[k(R_{v'}-A(k))]\kappa^{2};
\end{equation}
where $\kappa = {\bf d}\cdot \E/2\hbar$ is the Rabi frequency of the laser
of intensity
$I$,
$A(k)$ represents a shift from the free solution, and $\Delta_{v'}$
(again) is the detuning of the level $v'$ from the dissociation limit.
This expression is valid for $s$-wave scattering and at low detunings. In
the limit
$k\rightarrow 0$, $A(k_{R})\rightarrow a$, and
the maximum of ${\cal R}_{v'}$ is found to be at 
$\hbar\delta\simeq k_{B}T/2$, so that
\begin{equation}
   {\cal R}_{v'} \simeq \rho \lambda_{D}^{3}
   e^{-1/2} \kappa^{2}\sqrt{\frac{2\pi^2\mu^2}{3\hbar^3\Delta_{v'}}}
   \sqrt{k_{B}T}(R_{v'}-a)^{2} \; .\label{eq:pillet}
\end{equation}
This equation is similar to Eqs.(\ref{eq:rate})-(\ref{eq:dipole}).
Notice that since $\lambda_{D}\propto 1/\sqrt{T}$, 
the rate scales as $1/T$ and
$\Delta_{v'}^{-1/2}$ for small detuning. For lithium at 0.140 mK
(corresponding to the Doppler temperature),
assuming a laser intensity of 1000 W/cm$^{2}$ (so that
$\kappa=0.8\times 10^{10}$ s$^{-1}$: see \cite{pillet} for details),
and a detuning $\Delta_{v'}\sim 1$ cm$^{-1}\simeq 30$ GHz (or
$4.5\times 10^{-6}$ hartree), we have $R_{v'}= 135\,a_{0}$ and
$\lambda_{D}=1489\,a_{0}=7.875\times 10^{-6}$ cm. For a density
$\rho=10^{11}$ cm$^{-3}$, this gives $\rho\lambda_{D}^{3}=4.88\times
10^{-5}$ and, neglecting the scattering length $a$, one gets
${\cal R}_{v'}\sim 9.6\times 10^{4}$ s$^{-1}$. If we scale the rate
to $T=1$ mK, we get $1.3\times 10^{4}$ s$^{-1}$, less than twice
the value 7140 $s^{-1}$ obtained from Eq.~(\ref{eq:rate}) using the
calculated value of $\alpha^{PA}_{v'=86}$ given above.
Notice that many assumptions are made in these estimates: 
nonetheless, expression (\ref{eq:pillet}) gives good order of
magnitude for the photoassociation rate for small detuning and low
temperatures.

Using the Doppler temperature as a typical temperature, and averaging
the rate using a linewidth corresponding to 5 MHz \cite{pillet}, one
finds typical rates for small detunings by scaling the numbers of
Table~\ref{tab2} with the appropriate temperature, density, laser
intensity, and detunings, according to
\begin{equation}
    {{\cal R}_{v'}} \simeq \frac{I}{I_{0}} \frac{\rho}{\rho_{0}}
    \frac{T_{0}}{T} \sqrt{\frac{\Delta_{0}}{\Delta_{v'}}}\;
    \overline{\cal R} \; .
\end{equation}
Here, $I_{0}=1000$ W/cm$^{2}$, $\rho_{0}=10^{11}$ cm$^{-3}$,
$\Delta_{0}=1$ cm$^{-1}$, and $T_{0}$ and $\bar{\cal R}$
are listed in Table~\ref{tab2}. Correspondingly, $\alpha^{PA}$ scales as
\begin{equation}
    \alpha_{v'}^{PA} \simeq \frac{T_{0}}{T}
\sqrt{\frac{\Delta_{0}}{\Delta_{v'}}} \;
                  \overline{\alpha}^{PA} \; .
\end{equation}
The characteristic intensities should then scale with the square root of
the detuning $\Delta_{v'}$, so that we have
\beq
I_{v'} = \sqrt{\frac{\Delta_{v'}}{\Delta_{0}}}\,I_0\,.
\label{DETSCA}
\eeq

  \begin{table}[b]
  \begin{center}
  \caption{\protect\narrowtext
  Approximate scaled photoabsorption rates $\overline{\cal R}$
  and rate coefficients $\overline{\alpha}^{PA}$ for weakly bound
  molecular states. The values of $\overline{\cal R}$, and $T_{0}$ are from
  Pillet {\it et al.}~\protect\cite{pillet}, $I_{0}=1000$ W cm$^{-2}$,
  $\rho_{0}=10^{11}$ cm $^{-3}$, and $\Delta_{0}=1$ cm$^{-1}$.
  }
  \begin{footnotesize}
  \begin{tabular}{cccccc} \\
   Atom & $T_{0}$ & $\overline{\cal R}$ & $\lambda$
        & $\varphi$ & $\overline{\alpha}^{PA}$ \\
        & mK    & $10^{4}$ s$^{-1}$ & nm
        & $10^{21}$ cm$^{-2}$ & cm$^{5}$ \\ \hline
    &    &    &    &   &    \\
  Li & 0.140 & 45 & 671 & 3.4 & $1.3\times 10^{-27}$ \\
  Na & 0.240 & 22 & 589 & 3.0 & $7.3\times 10^{-28}$ \\
  K  & 0.140 & 25 & 766 & 3.9 & $6.6\times 10^{-28}$ \\
  Rb & 0.140 & 13 & 780 & 3.9 & $3.3\times 10^{-28}$ \\
  Cs & 0.125 & 10 & 852 & 4.3 & $2.3\times 10^{-28}$ \\
   &    &    &    &   &  \\
  \end{tabular}
  \end{footnotesize}
  \label{tab2}
  \end{center}
  \end{table}

The values of $\lambda$, $\varphi$ and $\overline{\alpha}^{PA}$ are
listed in Table~\ref{tab2}. To compare the rates for the various alkali
metals, it is convenient to express them for the same parameters. Assuming
$I_0=1000$~W/cm$^{2}$, $\rho=10^{11}$ cm$^{-3}$, $T=100$ $\mu$K, and
$\Delta_0=1$ cm$^{-1}$, we obtain the values listed in Table~\ref{tab3}.
The photoassociation rates vary between $1.2\times 10^{5}$ s$^{-1}$
(or $\alpha^{PA}\sim 2.9\times 10^{-28}$ cm$^{5}$) for Cs
and $6.3\times 10^{5}$ s$^{-1}$
(or $\alpha^{PA} \sim 1.8\times 10^{-27}$ cm$^{5}$)
for Li. From the expressions for ${\cal R}_{v'}$, one expects the rates
to scale like $1/2\mu$. If we were to multiply the rates by $M$, the mass
number, we notice that, except for Li, the scaled rates indeed are
similar. The variations left are due to slightly different Rabi frequencies
\cite{pillet}.
  \begin{table}[t]
  \begin{center}
  \caption{\protect\narrowtext
  Approximate photoassociation rates $\cal R$, rate coefficients $\alpha^{PA}$,
  and characteristic intensities $I_{v'}$ for the alkalis.
  The fixed parameters are $I_{0}=1000$ W/cm$^{2}$,
  $\rho_{0}=10^{11}$~cm$^{-3}$, $\Delta_{0}=1$ cm$^{-1}$, and T=100
$\mu$K,.
Also listed are the reference densities
$\lambdabar^{-3}$ (in cm${}^{-3}$) and frequencies $\epsilon_R$ (in
$2\pi\,{\rm kHz}$).}
  \begin{footnotesize}
  \begin{tabular}{cccccc} \\
   Atom & ${\cal R}_{v'}$ & $\alpha^{PA}_{v'}$ & $I_{v'}$
&$\lambdabar^{-3}$ &
$\epsilon_R$ \\
        & $10^{4}$s$^{-1}$& cm$^{5}$ & ${\rm mW\,cm}$${}^{-2}$ &
cm${}^{-3}$&
$2\pi\,{\rm kHz}$\\
\hline
        & & & &  \\
  ${}^7$Li &63& $1.8\times 10^{-27}$ & 3.7 & $8.21\times 10^{14}$   &
63.3\\
  ${}^{23}$Na &53& $1.7\times 10^{-27}$ &0.47 & $1.21\times 10^{15}$  &
25.0\\
  ${}^{39}$K  &35& $9.2\times 10^{-28}$ & 0.18& $5.52\times 10^{14}$  &
8.72\\
  ${}^{87}$Rb &18& $4.6\times 10^{-28}$ & 0.069 & $5.23\times 10^{14}$  &
3.77\\
  ${}^{133}$Cs &12& $2.9\times 10^{-28}$ &0.039 & $4.01\times 10^{14}$ &
2.07\\
     & & & &\\
  \end{tabular}
  \end{footnotesize}
  \label{tab3}
  \end{center}
  \end{table}

Finally, one has to be careful when using the values of
Table~\ref{tab3}, in which the effect of scattering lengths are not
taken into account. These effects can be significant, like in the case
of $^{6}$Li, $^{85}$Rb or $^{133}$Cs, where large negative
scattering lengths induce larger rates. Also, at larger
detunings, one probes deeper region of the excited electronic
state, and shorter distances of the lower state continuum
wave function, where the exact nodal structure will play
an important role (see \cite{COT95,cote-pra,cote-jms} and
Fig.~\ref{LIRATE}).

As already noted, the values for the saturation intensity in
Table~\ref{tab3} are to be construed as estimates only. Also, as they are,
they apply to a very weakly bound molecular state ($\Delta_{v'}=1~{\rm
cm}^{-1}$). Such high-lying states are likely to be easily perturbed by
atom-molecule collisions, so that in practice one might resort to more
bound molecular states. Nonetheless, these saturation intensities are
remarkably low, occasionally much lower than our initial generic estimate of
$10\, {\rm W cm}^{-2}$~\cite{JAV99,JAV99a}. From the experimental viewpoint
this is encouraging, as the requirements on laser intensity are greatly
moderated. The flip side is that the usable intensity appears to
be limited by rogue photoassociation; for densities such that
$\lambdabar^3\rho\sim1$, the maximum photoassociation Rabi
frequency is of the order of the recoil frequency $\epsilon_R$ . For
heavier alkalis the saturation intensities may indeed be low, but small
recoil frequency and slow coherent photoassociation are the prices to pay.

\subsection{Discussion of an experiment}\label{HEIZEXP}

In the experiment of Wynar {\it et al}., Ref.~\cite{WYN00}, the authors
studied photoassociation of a ${}^{87}$Rb condensate in a two-color Raman
configuration. From our standpoint, the reported data fall into two
categories. First, the width and shift of the two-photon resonance line was
studied as a function of the intensities of the two lasers. We are going
to use these measurements to estimate the bound-bound Rabi frequency.
Second, the authors measured the width and the shift of the same resonance
line as a function of the density of the condensate. At low intensities,
elastic atom-atom and atom-molecule collisions turn out to have a major
effect on the resonance parameters. However, these effects were
modeled quantitatively in Ref.~\cite{WYN00}, and one may then get at the
quantities pertaining to photoassociation. We use this type of data
to determine the characteristic frequency of coherent photoassociation.

\subsubsection{Adapting the theory}
We first cast our theoretical approach in a form that allows for direct
juxtapositions with the analysis of experimental data in Ref.~\cite{WYN00}.
We ignore elastic collisions because a comparison with Ref.~\cite{WYN00}
turns out to be possible even without considering them explicitly, and
inelastic collisions because their presence was not conclusively
established in Ref.~\cite{WYN00}.

We begin by adding the spontaneous decay of the intermediate state at the
rate $\Gamma_s$ as a nonhermitian term in the Hamiltonian for the $\Lambda$
system, Eq.~\eq{3MH}. This gives
\beq
{H\over\hbar} = \delta\,g^\dagger g + (\Delta+\delta)b^\dagger b +\ldots -
i{\Gamma_s\over2}\,b^\dagger b\,.
\label{3MHD}
\eeq
It is known in quantum optics that introducing an imaginary part to the
energy of the decaying state gives correct results as long as the unstable
state decays irreversibly only to states that are outside of the state
space included in the model.

We then repeat much of the analysis of Sec.~\ref{CONSGAMMA} for the
Hamiltonian~\eq{3MHD}, with a number of tricks and approximations. First, we
take the limit of large intermediate detuning,
$|\Delta|\gg\Gamma_s,|\delta|$. Second, we assume that the bound-bound Rabi
frequency $\chi$ is much larger than the free-bound Rabi frequency
$\Omega$. This implies that in the estimates of the line widths and decay
rates, such as those following from Eq.~\eq{IMFRAC}, only the
$\chi^2$ terms need be kept. Third, we ignore light-induced
atom-atom interactions. Fourth, we scale the the operators as before,
$a = \sqrt{N}\alpha$, etc., which brings out the photoassociation frequency
$\Omega$ explicitly. Fifth, after deriving the equations of motion for the
atomic operator $\alpha$ and final-state molecule operator $\gamma$ by
adiabatic elimination of the intermediate-state operate $\beta$, in the
spirit of semiclassical approximation we replace the operators $\alpha$,
$\beta$, and $\gamma$ with
$c$ numbers. At this stage we have, in analogy with Eqs.~\eq{HEQ1}
and~\eq{HEQ2}, the equations
\bea
\dot{\gamma} &=& -i \left(\delta-{\chi^2\over\Delta}\right)\gamma
- {\Gamma_s \chi^2\over 2 \Delta^2}\,\gamma
+\half i\,{\Omega\chi\over\Delta}\, \alpha^2,\\
\dot{\alpha} &=&i\,{\Omega\chi\over\Delta} \alpha^* \gamma\,.
\eea

To develop the rate approximation, we first note that the probability
that an atom belongs to the stable state $g$, $P_g = 2|\gamma|^2$, satisfies
the equation of motion
\beq
\dot{P}_g = -{\Gamma_s\chi^2\over\Delta^2}\,P_g +
i{\Omega\chi\over\Delta}(C-C^*)\,.
\label{GRMTN}
\eeq
The ``coherence'' $C=\alpha^2\gamma^*$ has the equation
of motion
\beq
\dot{C} = i{\Omega\chi\over\Delta}[P_g P_a-\half P_a^2] +
\left[
i\left(\delta-{\chi^2\over\Delta} \right)-{\Gamma_s\chi^2\over2\Delta^2}
\right] C\,,
\eeq
where, in turn $P_a = |\alpha|^2$ is the probability that an atom remains
in the system unassociated.

We could go on and derive a corresponding equation of motion for $P_a$.
However, here we assume that $P_a\sim1$, and that correspondingly
$P_g\ll1$. We solve for the coherence $C$ adiabatically by setting
$\dot{C}=0$, and insert the result into Eq.~\eq{GRMTN} to obtain a rate
equation for the population of the stable molecular state $g$,
\beq
\dot{P}_g = -{\Gamma_s\chi^2\over\Delta^2}\,P_g
+
{\left[{\Omega\chi\over\Delta}\right]^2
\left[{\Gamma_s\chi^2\over2\Delta^2} \right]
\over
\left[\delta-{{\chi^2\over\Delta}} \right]^2 +
\left[{\Gamma_s\chi^2\over2\Delta^2} \right]^2
}\,P_a^2\,.
\label{gRATE}
\eeq
This  displays a shift of the ground state by
\beq
\Delta\varepsilon_L = {\chi^2\over\Delta}\,,
\eeq
a linewidth
\beq
{\gamma_L\over2} = {\Gamma_s\chi^2\over2\Delta^2}\,,
\eeq
and a corresponding direct decay of bound molecules at the rate
$\gamma_L$.

With the assumption $P_g \ll 1$ we have made the rate equation~\eq{gRATE}
essentially one way, i.e., ignored transitions from molecules back to the
atoms. The semiclassical
approximation in general requires that there are many molecules in a
quantum state, so that the molecules make a condensate. However, we believe
that, by ignoring transitions from molecules back to atoms, we have
divorced the rate equation~\eq{gRATE} from any coherence requirement for
the molecules. As the experiments do not claim a molecular condensate,
this is essential for the analysis that follows.

\subsubsection{Comparison with experiments}

Wynar {\it et al}., Ref.~\cite{WYN00}, discuss the light shift of
the resonance line. In their case the frequencies of the two lasers are
quite close so that both of them cause line shifts and broadenings, but
this is included in their analysis. Translating back to our case with only
one laser frequency on the bound-bound transition, the comparison between
our formulation and Ref.~\cite{WYN00} reads
\beq
{\chi^2\over\Delta} = {\beta I_2\over\Delta_1}\,,
\eeq
where $I_2$ is the intensity (in W cm${}^{-2}$) of the bound-bound light,
$\Delta_1\equiv\Delta=2\pi\times150\,{\rm MHz}$ is the intermediate
detuning, and $\beta=3.5\times10^9\,{\rm m^2 W^{-1}s^{-2}}$ is a parameter
that we deduce from their Fig.~3a. This gives the approximate formula
\beq
\chi \simeq 0.94 \,\sqrt{I_2\over{\rm W cm^{-2}}}\times2\pi\,{\rm MHz}\,.
\eeq

Wynar {\it et al}., Ref.~\cite{WYN00}, also present an analysis of line
shapes at low intensity. The rate coefficient for two-photon
photoassociation on resonance, $K_0 = 9\times10^{-14}\,{\rm
cm^3\,s^{-1}}$, is one of the fitted parameters. The resonance
photoassociation rate from our Eq.~\eq{gRATE} and the corresponding
expression in Ref.~\cite{WYN00}, ignoring inelastic processes and with $n
\equiv\rho$, boil down to
\beq
{\Omega^2\over\Gamma_s} = {n K_0}\,.
\eeq
Given the spontaneous decay rate of this molecular level, $\Gamma_s =
12\times2\pi\,{\rm MHz}$~\cite{WYN00}, and noting that their cited value of
$K_0$  was for the intensity of the photoassociation laser $I_1=0.51\,{\rm
W\,cm^{-2}}$, we find the scaling formula
\beq
\Omega \simeq 5.8\,\sqrt{{\rho\over 10^{14}{\rm cm^{-3}}}\,{I_1\over{\rm
W cm^{-2}}}}\,\times 2\pi\,{\rm kHz}\,,
\eeq
which gives the characteristic intensity $I_{v'} = 0.08\,{\rm W cm^{-2}}$.

In the experiment, the binding energy of the primary photoassociated state
was 23~cm${}^{-1}$. In contrast, for $\Delta_{v'} = 1\,{\rm cm}^{-1}$,
Table~\ref{tab3} gives the characteristic intensity $0.07\,{\rm mW
cm^{-2}}$. The theoretical and experimental characteristic intensities
differ by three orders of magnitude, but they are also for
different binding energies. The difference in characteristics
intensities is no cause for concern, as we do not know at all
how the detuning 23~cm${}^{-1}$ for Rb is related to the  minima of the
photoassociation rates like those shown in Fig.~\ref{LIRATE} for Li.

\subsubsection{Reaching coherence}

Suppose that the density of the ${}^{87}$Rb gas equals
$\rho=\lambdabar^{-3}$, a factor of two larger than quoted in
Ref.~\cite{WYN00}, the laser intensities are $I_1=I_2=10\,{\rm
W\,cm^{-2}}$, typical values in photoassociation experiments, and the
intermediate detuning is as in Ref.~\cite{WYN00}. Then the two-photon Rabi
frequency becomes $\Omega\chi/\Delta = 1\,\times 2\pi\,{\rm kHz}$, while
the effective linewidth is $\chi^2\Gamma_s/2\Delta^2= 2\times 2\pi\,{\rm
kHz}$. For coherent phenomena we need a Rabi frequency at least comparable
to the linewidth, a condition that is not quite satisfied.
But if one were simply to take the intensities
$I_1=100\,{\rm W\,cm^{-2}}$ for the photoassociating light and
$I_2=1\,{\rm W\,cm^{-2}}$  for the bound-bound light, Rabi frequency
remains unchanged and linewidth goes down by a factor of 100. One is then
in principle deep in the regime of coherent photoassociation. Also, the
effective Rabi frequency would be smaller than the recoil frequency, so
that rogue photoassociation need not yet be a fatal problem.

On the other hand, eyeballing from Fig.~2 of Ref~\cite{WYN00}, the linewidth
due to elastic atom-atom collisions would be of the order of $5\times
2\pi\,{\rm kHz}$, so elastic collisions would interfere with coherent
photoassociation. Rabi oscillations would be seriously impeded, rapid
adiabatic passage maybe less. One could compensate by lowering the
density by two orders of magnitude and at the same time increasing the
intensity of the photoassociating laser by another two orders of magnitude
to keep the effective Rabi frequency constant, but with lowering of the
density one would  wind up in trouble with rogue photoassociation.

Experimentally, one can think of varying the intensity of the two lasers,
the intermediate state, the intermediate detuning, and the density of the
sample. The conditions one must watch out for are that effective Rabi
frequency be larger than the effective damping rate and line broadening due
to elastic (and inelastic) collisions, yet not so large as to cause
excessive rogue photoassociation. If the parameters could be varied freely,
within our physical model a solution is always possible. Unfortunately, one
must live with some practical limitations on, say, laser intensities. The
present experiment of Ref.~\cite{WYN00} is tantalizingly close to coherent
photoassociation, but to really get there may require a careful
optimization of the experimental parameters and a solid understanding of
what is technically feasible in a given experiment. We will not attempt to
enter such discussions.

\section{Concluding remarks}\label{CONCL}

We have developed theory for coherent photoassociation of a
Bose-Einstein condensate of atoms, which typically leads to a Bose-Einstein
condensate of molecules. Phenomena analogous to coherent optical transients
in few-level systems, such as Rabi flopping and rapid adiabatic passage
between atomic and molecular condensates are expected. Coherent
phenomena depend on the Bose enhancement of the dipole moment matrix
element, and cannot occur in a nondegenerate gas, at least not in the
thermodynamic limit. Bose-Einstein condensation and photoassociation are
both currently front-line research and the marriage thereof might still be a
technological challenge, but coherent optical transients in
photoassociation should eventually be feasible in experiments.

\acknowledgments
This work is supported in part by the NSF, Grant No. PHY-9801888, and
by NASA, Grant No. NAG8-1428. Additional support was provided by the
Connecticut Space Grant College Consortium. One of us [JJ] thanks
S\'{a}ndor Varr\'{o} for interesting discussions; in particular, for
pointing out Ref.~\cite{VOLREF}.

\appendix
\section{Dipole matrix element}\label{DMA}

In our bare-bones example we consider the situation in which we either have
one molecule at rest and no atoms, or no molecule and two atoms in states
$\kvec$ and $-\kvec$. Then the relative momentum becomes $\hbar\kvec$.
This peculiarity, and also the reason for the factor $\half$ in
Eq.~\eq{ID}, is because the relative momentum for two entities with
momenta
$\p_1$ and $\p_2$ should be defined as $\half(\p_1-\p_2)$ to make it the
conjugate of the usual relative position
$\rvec_1-\rvec_2$. We thus have the possible state vectors
\beq
|\psi\rangle = \left(\beta\,b^\dagger + \sum_\kvec\alpha_\kvec a^\dagger_\kvec
a^\dagger_{-\kvec}\right)|0\rangle\,,
\eeq
where $\beta$ and $\alpha_k$ are complex amplitudes to be determined, and
$|0\rangle$ stands for the vacuum of atoms and molecules. The states for
$\alpha_\kvec$ and $\alpha_{-\kvec}$ are the same, so the sum over $\kvec$ runs
only over half of the possible $\kvec$ values; say, those with $k_x>0$.

Following Ref.~\cite{JAV99}, we write the time dependent Schr\"odinger
equation from the Hamiltonian~\eq{HK} in terms of the coefficients $\beta$
and $\alpha_k$, assuming a plane wave of light. In order to avoid an
inessential complication, we ignore photon recoil and simply set~$\q=0$.
The result is
\bml
\bea
\dot\alpha_\kvec &=& -i\left({\hbar \kvec^2\over 2\mu} -\delta_0\right)\alpha_\kvec
+ i {\db^*\cdot\E^*\over 2\hbar}\,\beta,\\
\dot\beta &=& i\sum_\kvec{\db\cdot\E\over 2\hbar}\,\alpha_\kvec\,.
\eea
\eml
The QC approach discussed in Refs.~\cite{JAV98}
and~\cite{MAC99} displays precisely the same structure, evidently
identifying $\db\cdot\E/2\hbar$ as the QC Rabi frequency $\kappa$ of
those papers.

There is a catch, however. Because of the Bose-Einstein statistics,
states with the relative momenta for two atoms $\kvec$ and $-\kvec$ are the same,
so the density of states for the relative motion  is half of what it was
for Maxwell-Boltzmann atoms. To obtain the same photodissociation 
rate, we therefore have
to make the matrix element a factor of
$\sqrt{2}$ larger than in the case of distinguishable atoms. This is the
$\sqrt{2}$ in Eq.~\eq{ID}. The rate of photodissociation is
proportional to the square of the matrix element and picks up a factor of
two, which compensates for the missing half of phase space density.

The second-quantized Hamiltonian is chosen so that it gives the right
photodissociation rate. What we did not fully realize in Ref.~\cite{JAV99}
is that the same factor of $\sqrt{2}$ also leads to {\em twice\/} the
photoassociation rate that one would obtain for Maxwell-Boltzmann atoms,
given the Franck-Condon factor from the standard calculations or as deduced
from the photodissociation rate. It appears that in the literature 
the rate for $s$-wave
photoassociation is usually calculated low by a factor of two. 
Besides, by a simple
extension of the statistics arguments, $p$-wave photoassociation should
vanish for bosons. Are these problems?

We think not. In photoassociation experiments no particular care is 
usually taken to
polarize the atoms. They are not all in the same internal state, and
therefore not all photoassociation events are between indistinguishable
atoms. This should reduce the factor-of-two boson enhancement of the 
photoassociation rate
for even partial waves, and call forth odd partial waves. The remaining
discrepancies with the conventional calculations may well be within
experimental and theoretical uncertainties.

It would be of some interest to investigate photoassociation of a
polarized low-temperature gas experimentally. We believe that the statistics
effects are real, and should be observable. After all, few colleagues seem
to have a problem with the assertion that there are no $s$-wave collisions
for polarized fermions. Photoassociation experiments in a polarized gas
should make an interesting test of our phenomenological Hamiltonian.

\section{Kernel for photoassociation in field theory}\label{PAK}

The process we followed in the main text in deriving the integral kernel
$\db(\rvec)$ [as in, e.g., Eq.~\eq{KERFOR}] was a phenomenological mix
between formally pure field theory and collisional physics.

In fact, to obtain the dipole matrix elements $\db_{\kvec\kvec'}$ in
Eq.~\eq{HK}, we carry out an integral of the form
\beq
\db_{\kvec\kvec'} \equiv \db(\kvec-\kvec') =\db\int
d^3r\,\bar\psi^*(\rvec)
\bar\phi_{\kvec-\kvec'}(\rvec)\,,
\label{DMELE}
\eeq
where $\bar\psi$ is the bound-state molecular wave function and
$\bar\phi_{\kvec-\kvec'}$ describes the state of two atoms with the relative
momentum $\hbar(\kvec-\kvec')$. Now,
in the transformation to field theory~\eq{TOFIELDS} we used plane waves
as the wave functions for states with a given wave vector. In
the matrix element~\eq{DMELE} we should correspondingly use plane waves
to describe the relative motion, as in
\beq
\bar\phi_{\kvec-\kvec'}(\rvec) =
{1\over\sqrt{V}}\,e^{i(\kvec-\kvec')\cdot\rvec}\,.
\eeq
However, in our derivation we did not not use plane waves, but distorted
plane waves that take into account atom-atom interactions, and
only asymptotically (at large distances) corresponds to the relative
momentum $\kvec-\kvec'$.

This makes a difference. For instance, if the matrix element~\eq{DMELE} is
computed using pure plane waves, the final photoassociation kernel reads
\beq
\db(\rvec) = \sqrt{2}\,\db\,\bar\psi^*(r)\,,
\eeq
instead of something akin to~\eq{DR1}. Depending on the value of the
scattering length, there could be a substantial difference in the overall
numerical factor also in the contact-interaction form for atom-molecule
coupling, Eq.~\eq{CIF}. The question is, which method is ``correct'';
pure or distorted plane waves?

First of all, atoms {\em do\/} interact, which in general {\em will\/} have
an effect on photoassociation. One who wishes to analyze 
photoassociation quantitatively will have to
take atom-atom interactions into account. A field theorist pursuing pure
plane waves would have to consider atom-atom interactions explicitly, and
work out the consequences in photoassociation {\it ab initio}. This is 
probably a major chore.
Our hope is that, with our distorted-plane wave trickery, we have
phenomenologically captured most of the effect of atom-atom interactions on
photoassociation.

On the other hand, if we use the distorted plane waves to calculate the
photoassociation matrix element and {\em in addition} introduce a 
contact interaction
model for atoms as in~\eq{AAI}, the suspicion arises that
we are double-counting atom-atom interactions. Further approximations, such
as the contact-interaction form for photoassociation or classical 
field theory, could
either exacerbate or ameliorate the double-counting.

Disentangling atom-atom interactions and photoassociation might make 
an interesting
theoretical exercise, but we do not attempt it in this paper.
Instead, we proceed with a model that allows for both
distorted-plane wave matrix elements for photoassociation, and an 
explicit atom-atom
interaction. In this way, we at least get the
limiting cases of photoassociation in a dilute gas and atom-atom 
interactions in the
absence of photoassociation basically right with little effor


%
%

%
%

\begin{references}
\bibitem{note-rate} The rate coefficients listed in
Refs.~\protect\cite{cote-pra,cote-jms} must be
   multiplied by $(2\pi )^{5}$, and we do not assume any hyperfine
   mixing (neglect $\omega_{0}$ in~\protect\cite{cote-pra,cote-jms}).
\bibitem{note-hulet}
   We determined the binding energy of level $v'=58$ of $^{7}$Li triplet
   state to be $3666\pm 2$ GHz by fitting the known levels, and it was
   recently measured by R.G. Hulet and co-workers to be 3667 GHz.
\end{references}
\end{document}